\documentclass[aps,prd,onecolumn,amsmath,amssymb,showkeys,nofootinbib,floatfix,superscriptaddress,]{revtex4-2}
\usepackage[colorlinks=true,
            linkcolor=blue,
            citecolor=blue,
            urlcolor=blue]{hyperref}
\usepackage{xcolor}
\usepackage{amsmath}
\usepackage{amssymb}
\usepackage{physics}
\usepackage{graphicx}
\usepackage{svg}
\usepackage{url}
\usepackage{ulem} 
\usepackage{multirow}
\usepackage{listings}
\usepackage{float}
\restylefloat{table}
\usepackage{longtable}
\usepackage{orcidlink}
\usepackage{rotating}
            
\newcommand{\msol}{\Delta m_{\text{sol}}}
\newcommand{\matm}{\Delta m_{\text{atm}}}
\newcommand{\mso}{\Delta m_{s_1}}
\newcommand{\mst}{\Delta m_{s_2}}

\newcommand{\nubb}{$0\nu\beta\beta$}
\newcommand{\eV}{\text{eV} }
\lstdefinestyle{PythonStyle}{
    language=Python,
    basicstyle=\ttfamily\footnotesize,
    keywordstyle=\color{blue}\bfseries,
    commentstyle=\color{gray},
    stringstyle=\color{red},
    numbers=left,
    numberstyle=\tiny,
    stepnumber=1,
    numbersep=5pt,
    backgroundcolor=\color{lightgray!20},
    frame=single,
    captionpos=b,
    breaklines=true,
    showstringspaces=false,
    tabsize=4
}

\begin{document}
\title{Neutrino mass variables in 3 active and 2 sterile neutrino scenario}

\author{Srubabati Goswami$^{}{}$ \orcidlink{0000-0002-5614-4092}}
\email{sruba@prl.res.in}
\affiliation{\vspace{0.2cm}$^{}$Theoretical Physics Division, Physical Research Laboratory, Ahmedabad - 380009, India}
\author{Hemanth M.$^{}{}$\orcidlink{0009-0000-3254-713X}} 
\email{hemanth.m@ug.cusat.ac.in}
\affiliation{\vspace{0.2cm}$^{}$Cochin University of Science and Technology, Kalamassery, Kochi - 682022, India}
\author{Debashis Pachhar$^{}{}$ \orcidlink{0000-0001-8931-5321}}
\email{debashispachhar@prl.res.in }
\affiliation{\vspace{0.2cm}$^{}$Theoretical Physics Division, Physical Research Laboratory, Ahmedabad - 380009, India}
\author{N Rajeev$^{}{}$ \orcidlink{0000-0003-2805-8178}}
\email{rajeevneutrino@gmail.com}
\affiliation{\vspace{0.2cm}$^{}$Theoretical Physics Division, Physical Research Laboratory, Ahmedabad - 380009, India}
\affiliation{\vspace{0.2cm}$^{}$Joint Institute for Nuclear Research, Dubna, Moscow Region, 141980, Russia}

\pacs{}
\begin{abstract}
    The three-flavor framework of neutrino oscillations successfully explains most experimental results, but persistent anomalies at short- and long-baseline experiments hint at the existence of additional light sterile states. In particular, eV-scale sterile neutrinos are motivated by LSND and MiniBooNE results, while sub-eV sterile states with mass-squared differences at the $10^{-2}$ and $10^{-5}$~eV$^2$ scales have been proposed to address the T2K--NO$\nu$A tension and the absence of the expected  upturn in the solar neutrino energy  spectrum, respectively. Such sterile states are singlets under the Standard Model gauge group and mix only through their admixture with active neutrinos. In this work, we investigate the phenomenology of the $3+2$ scenario, incorporating one eV-scale sterile neutrino together with a sub-eV state, and analyze their impact on absolute-mass related observables: the sum of neutrino masses $\Sigma$ constrained by cosmology, the effective electron neutrino mass $m_\beta$ from $\beta$ decay, and the effective Majorana mass $m_{\beta\beta}$ probed in neutrinoless double $\beta$ decay. We demonstrate that the presence of two sterile states can significantly modify the allowed parameter space compared to the three-flavor and $3+1$ frameworks, with some mass-ordering schemes already disfavored by current cosmological and laboratory limits. Finally, we assess the implications of upcoming sensitivities from KATRIN, Project~8, and LEGEND-1000, highlighting the complementary role of sub-eV sterile neutrinos in probing physics beyond the minimal three-flavor paradigm.
\end{abstract}

\maketitle

\section{Introduction}

The Standard Model (SM) of particle physics has provided an extraordinarily successful description of fundamental particles and their interactions, confirmed by precision measurements at colliders and fixed-target experiments~\cite{Glashow:1961tr,Weinberg:1967tq}. Nevertheless, the SM is incomplete: it offers no viable dark matter candidate, does not explain the baryon asymmetry of the universe, and originally assumed neutrinos to be massless with independently conserved lepton family numbers~\cite{Weinberg:1979sa}. This paradigm was overturned by the discovery of neutrino oscillations. Compelling evidence first came from atmospheric neutrino observations at Super-Kamiokande~\cite{Super-Kamiokande:1998kpq} and solar neutrino measurements at SNO~\cite{SNO:2002tuh}, subsequently confirmed by accelerator~\cite{K2K:2004iot,MINOS:2020llm,T2K:2011qtm,NOvA:2021nfi} and 
reactor experiments~\cite{KamLAND:2013rgu,DoubleChooz:2012gmf,RENO:2012mkc,DayaBay:2013yxg}. These observations firmly established that the three active neutrinos mix via the PMNS matrix, characterized by two independent mass-squared splittings $(\Delta m^2_{21},\Delta m^2_{31}; ~\Delta m_{ij}^2 = m_i^2-m_j^2)$, three mixing angles~$(\theta_{12}, \theta_{13}, \theta_{23})$, and one CP-violating phase~$(\delta_{CP})$. 

While the three-flavor framework accounts for the bulk of oscillation data, several short-baseline anomalies remain unresolved. The LSND experiment reported a $\bar{\nu}_e$ excess in a $\bar{\nu}_\mu$ 
beam~\cite{LSND:2001aii}, later corroborated by MiniBooNE~\cite{MiniBooNE:2020pnu}, while Gallium source experiments (GALLEX~\cite{GALLEX:1997lja}, SAGE~\cite{Abdurashitov:1996dp}, BEST~\cite{Barinov:2021asz}) observed deficits in $\nu_e$ event rates. These anomalies can be addressed by introducing at least one additional neutrino state at $\mathcal{O}(1)$~eV~\cite{Goswami:1995yq,Gomez-Cadenas:1995epo}. Since measurements of the invisible $Z$-boson width at LEP constrain the number of light active neutrinos to three~\cite{CMS:2022ett}, any additional light neutrino state must be sterile, i.e., singlet under the SM gauge group. This has motivated extensive studies of scenarios containing three active and $N$ sterile neutrinos. The simplest realization is the $3+1$ framework, which introduces one additional sterile state associated with an extra mass-squared splitting $\Delta m^2_{41}\sim\mathcal{O}(1)~\mathrm{eV}^2$.

Early global fits showed that $3+1$ models only partially accommodate the LSND and MiniBooNE anomalies, suffering from tensions between appearance and disappearance datasets~\cite{Giunti:2011gz,Giunti:2011hn,Giunti:2011cp}. Extended $3+2$ scenarios, which introduce a second mass-squared splitting $\Delta m^2_{51}$ and additional CP-violating phases, can improve the 
overall fit quality. Reanalyses incorporating NEOS~\cite{NEOS:2016wee}, DANSS~\cite{Alekseev:2016llm}, and IceCube~\cite{IceCube:2016rnb} data have further shifted or excluded allowed sterile parameter 
regions~\cite{Gariazzo:2017fdh}, and extensions involving non-standard interactions or non-unitary mixing broaden the landscape further~\cite{Fong:2016yyh,Denton:2018dqq,Giunti:2019hkv,Minakata:2025azk}. Recent constraints come from ANTARES atmospheric neutrinos~\cite{ANTARES:2018rtf}, KATRIN kinematic searches~\cite{Giunti:2019fcj,Adams:2020nue,KATRIN:2020dpx,Goswami:2024ahm}, and reactor analyses~\cite{Serebrov:2021zuh,Serebrov:2021ndf}, while the latest MicroBooNE results find no evidence for $3+1$ oscillations~\cite{MicroBooNE:2025nll}.

A persistent tension exists between eV-scale sterile neutrinos favored by short-baseline data and cosmological observations. Such states typically thermalize in the early Universe, yielding $N_{\rm eff}\simeq 4$, in conflict with CMB and large-scale structure data~\cite{Archidiacono:2012ri,Gariazzo:2019gyi}. However, these bounds can be relaxed if sterile neutrinos are non-thermally produced through new sterile-sector interactions, modified expansion histories, or low reheating temperatures~\cite{Archidiacono:2014nda,Gelmini:2008fq}, allowing partial thermalization consistent with both laboratory and cosmological constraints.

Neutrinoless double $\beta$ decay~(\nubb) provides an independent probe of sterile neutrino scenarios, since active--sterile mixing induces additional contributions to the effective Majorana mass. Bounds on the $\sin\theta_{14}$--$m_{\rm lightest}$ plane from GERDA and KamLAND-Zen were derived in Ref.~\cite{Chakraborty:2019rjc}, and updated analytical constraints in the $\Delta m^2_{41}$--$\sin^2 2\theta_{14}$ plane using the latest KamLAND-Zen data were obtained in Ref.~\cite{Jana:2024xmc}, 
excluding substantial parameter space favored by BEST and Neutrino-4. The impact of a light sterile neutrino on the three neutrino mass observables: the cosmological sum, the kinematic $\beta$-decay mass, and the effective Majorana mass has been studied within $3+1$ and $3+2$ frameworks in Refs.~\cite{Goswami:2005ng,Goswami:2007kv,Goswami:2024ahm}.

Extensions to sub-eV sterile neutrinos have also been investigated. States with $\Delta m^2_{01}\sim(0.7\text{--}2)\times10^{-5}~\mathrm{eV}^2$ can suppress the low-energy solar $\nu_e$ survival probability~\cite{deHolanda:2010am,deHolanda:2003tx}, and including a very light sterile state can alleviate the T2K--NO$\nu$A tension~\cite{deGouvea:2022kma}. Implications for future detectors have been studied for DUNE~\cite{Chatterjee:2023qyr}, Hyper-Kamiokande, 
ESS$\nu$SB~\cite{Cabrera:2025fmj,Cabrera:2025qcs,KumarAgarwalla:2019blx,Agarwalla:2018nlx}, and IceCube-Upgrade~\cite{Cabrera:2025fmj}.

In this work, we investigate a $3+2$ scenario containing one eV-scale sterile neutrino, motivated by short-baseline anomalies, and one sub-eV sterile state, motivated by long-baseline and solar neutrino tensions. We focus on the impact of these sterile states on the absolute neutrino mass observables: the sum of neutrino masses from cosmology, the effective electron neutrino mass in $\beta$ decay, and the effective Majorana mass in neutrinoless double $\beta$ decay. We classify the possible mass spectra into four distinct ordering schemes, derive explicit expressions for these observables, and delineate the viable parameter space using the latest global oscillation fits, with implications for KATRIN, Project~8, and LEGEND-1000.

The paper is organized as follows. In Sec.~\ref{sec:neutrino_mixing} we present the $3+2$ neutrino mixing framework. Sec.~\ref{sec:classification} classifies the possible mass orderings. Sec.~\ref{sec:bound_3+2} discusses constraints from cosmology, $\beta$ decay, and neutrinoless double $\beta$ decay. Numerical results are given in Sec.~\ref{sec:results}, and Sec.~\ref{sec:summary} summarizes our conclusions.

\section{Neutrino Mixing Framework}
\label{sec:neutrino_mixing}
\subsection{The Standard Three-Neutrino Paradigm}
In the three-neutrino paradigm, the flavor eigenstates $\nu_\alpha = (\nu_e, \nu_\mu, \nu_\tau)^T$ are related to the mass eigenstates $\nu_i$ ($i=1,2,3$) by a unitary mixing matrix, the Pontecorvo–Maki–Nakagawa–Sakata (PMNS) matrix such that $ \nu_\alpha = \sum_{i=1}^{3} U_{\alpha i} \, \nu_i^{\rm mass}$. The PMNS matrix can be parameterized in terms of three mixing angles $(\theta_{12}, \theta_{23}, \theta_{13})$, one Dirac CP-violating phase $\delta$, and, if neutrinos are Majorana fermions, two additional phases $(\alpha_{21}, \alpha_{31})$ come into the picture. Its standard parameterization of the $3\times3$ PMNS matrix is given by,

\begin{equation}
U_{\rm PMNS} =
\begin{pmatrix}
c_{12} c_{13}                                          & s_{12} c_{13}                                         & s_{13} e^{-i\delta} \\
- s_{12} c_{23} - c_{12} s_{23} s_{13} e^{i\delta}    & c_{12} c_{23} - s_{12} s_{23} s_{13} e^{i\delta}     & s_{23} c_{13} \\
s_{12} s_{23} - c_{12} c_{23} s_{13} e^{i\delta}     & - c_{12} s_{23} - s_{12} c_{23} s_{13} e^{i\delta}   & c_{23} c_{13}
\end{pmatrix}
\cdot
\mathrm{diag}\,\!\big(1,\; e^{i\alpha_{21}/2},\; e^{i\alpha_{31}/2}\big),
\end{equation}

where \(c_{ij}\equiv\cos\theta_{ij}\), \(s_{ij}\equiv\sin\theta_{ij}\), \(\delta\) is the Dirac CP phase, and \(\alpha_{21},\alpha_{31}\) are the Majorana phases.

\subsection{The $3+2$ Sterile Neutrino Framework}

The $3+2$ framework extends the three-neutrino paradigm by introducing two additional sterile states, leading to five mass eigenstates in total. The flavor eigenstates can be written ,
\begin{equation}
\nu_\alpha = \sum_{i=1}^{5} U_{\alpha i} \, \nu_i^{\rm mass}, \qquad
\alpha = e, \mu, \tau, s_1, s_2,
\end{equation}
where $U$ is now a $5\times 5$ unitary mixing matrix, and $(s_1, s_2)$ denote the sterile flavors. The matrix $U_{5\times 5}$ can be parameterized as a sequence of complex
rotations in the $(i,j)$ planes, generalizing the standard PMNS structure~\cite{Cabrera:2024rgi} as,
\begin{eqnarray}
U_{5\times 5} &=&
R_{45}(\theta_{45})
R_{35}(\theta_{35},\delta_{35})
R_{25}(\theta_{25},\delta_{25})
R_{15}(\theta_{15},\delta_{15})\\ && \nonumber
R_{34}(\theta_{34})
R_{24}(\theta_{24},\delta_{24})
R_{14}(\theta_{14},\delta_{14})
\cdot \widetilde U_{\rm PMNS},
\end{eqnarray}
where $R_{ij}(\theta_{ij},\delta_{ij})$ denotes a complex rotation in the $(i,j)$ plane by angle $\theta_{ij}$ and phase $\delta_{ij}$, and $\widetilde U_{\rm PMNS}$ is the $5\times5$ matrix where the standard PMNS is embedded into the upper-left block and the lower-right block contains the extra $2\times2$ Majorana Phase matrix as,
\begin{equation}
\widetilde U_{\rm PMNS} =
\begin{pmatrix}
U_{\rm PMNS} & \mathbf{0}_{3\times2} \\
\mathbf{0}_{2\times3} & \mathbb{P}_{2\times2}
\end{pmatrix}, \label{eq:Upmns32}
\end{equation} 
where $\mathbb{P} = \text{diag}~\left(e^{i\alpha_{41}/2} , e^{i\alpha_{51}/2} \right)$. In total, the addition of two sterile states introduces 7 new mixing angles, 5 new Dirac CP phases, and 2 Majorana phases.   
Explicitly, the full $5\times 5$ mixing matrix takes the schematic form as,
\begin{equation}
U_{5\times 5} =
\begin{pmatrix}
U_{e1} & U_{e2} & U_{e3} & U_{e4} & U_{e5} \\
U_{\mu 1} & U_{\mu 2} & U_{\mu 3} & U_{\mu 4} & U_{\mu 5} \\
U_{\tau 1} & U_{\tau 2} & U_{\tau 3} & U_{\tau 4} & U_{\tau 5} \\
U_{s_1 1} & U_{s_1 2} & U_{s_1 3} & U_{s_1 4} & U_{s_1 5} \\
U_{s_2 1} & U_{s_2 2} & U_{s_2 3} & U_{s_2 4} & U_{s_2 5}
\end{pmatrix}, \label{eq:U5}
\end{equation}
where, the $(3\times 3)$ block $(U_{\alpha i}, \, \alpha=e,\mu,\tau,\, i=1,2,3)$ corresponds to the usual PMNS matrix. The entries $(U_{\alpha 4}, U_{\alpha 5})$ with $\alpha = e,\mu,\tau$
describe the active–sterile mixing elements controlled by $\theta_{i4}$ and $\theta_{i5}$. The $(2\times 2)$ bottom-right block $(U_{s_i j},\, i=1,2,\, j=4,5)$ governs the mixing between sterile states. 

\section{Classification of mass orderings in $3+2$ sterile neutrino framework}\label{sec:classification}
\begin{figure}[t]
\centering
\includegraphics[scale =0.5]{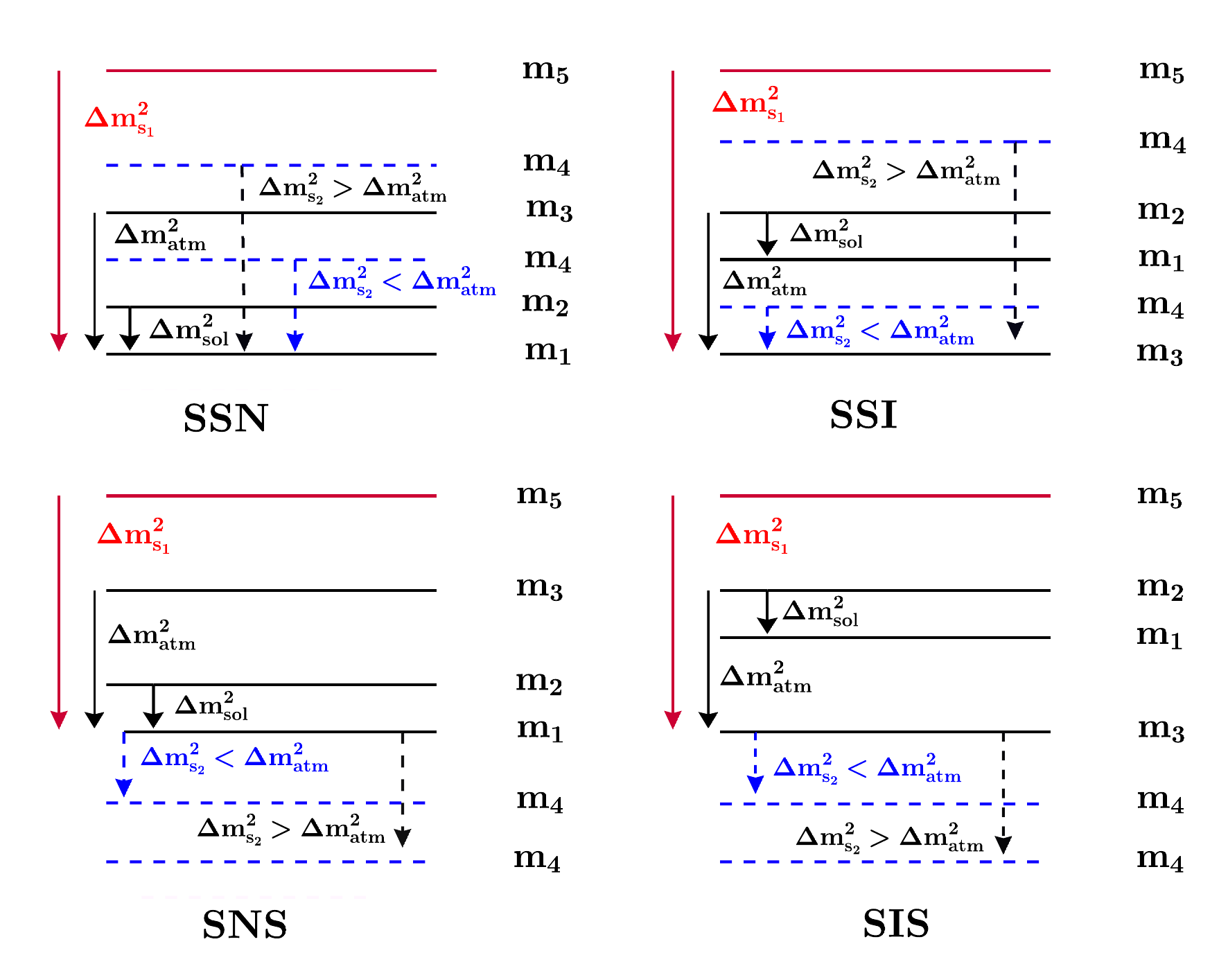}
\caption{Four mass spectra with three active and two sterile neutrinos. In the two cases above the lightest mass state is an active neutrino state and in the below the lightest state is a sterile state.}
\label{fig:3+2spectrum}
\end{figure}

In the standard three-neutrino paradigm the oscillation experiments are sensitive to the mixing angles and the {two independent mass-squared differences such as $\Delta m_{\text{sol}}^2 = \Delta m^2_{21} \equiv m_2^2 - m_1^2$, and $\matm^2~\equiv \Delta m^2_{3\ell} = m_3^2 - m_\ell^2$, where $\ell = 1 (2)$ for Normal Ordering (Inverted Ordering) of neutrinos. In the normal ordering (NO), we have $m_1 < m_2 < m_3$, with $\Delta m^2_{31} > 0$, whereas, in inverted ordering (IO) it is $m_3 < m_1 < m_2$, with $\Delta m^2_{32} < 0$.} In the same spirit, for the $3+2$ sterile neutrino framework, the presence of two sterile states gives rise to two new mass-squared differences which is defined as,
\begin{equation}
\Delta m_{s_1}^2 \equiv \Delta m^2_{5\ell} = m_5^2 - m_\ell^2, \qquad \Delta m_{s_2}^2 \equiv \Delta m^2_{4\ell} = m_4^2 - m_\ell^2, 
\end{equation}
where $\ell = 1 (3)$ for NO (IO) of active neutrino states. The presence of extra sterile neutrinos allow us to define different mass ordering schemes. There are four mass orderings that are possible: \textbf{SSN} (Sterile-Sterile-Normal hierarchy), \textbf{SSI} (Sterile-Sterile-Inverted hierarchy), \textbf{SNS} (Sterile-Normal-Sterile hierarchy), and the \textbf{SIS} (Sterile-Inverted-Sterile hierarchy) and shown in Fig.~\ref{fig:3+2spectrum}. 

\begin{itemize}
\item Sterile-Sterile-Normal Hierarchy Scheme (SSN): 
In this scheme, $m_{5}>m_{3}>m_{4}>m_{2}>m_{1}$ for ($\Delta m^{2}_{s_2}<\matm^2$) and $m_{5}>m_{4}>m_{3}>m_{2}>m_{1}$ for ($\Delta m^{2}_{s_2}>\matm^2$)  .The mass-squared difference of the heaviest neutrino mass state and the lightest neutrino mass state is given by $\Delta m^{2}_{s_1}\equiv m^{2}_{5}-m^{2}_{1}=1.3$ eV$^{2}$.
The mass-squared difference of the second sterile neutrino mass state with the lightest mass is varied as $\Delta m^{2}_{s_2}\equiv m^{2}_{4}-m^{2}_{1}=(10^{-4}-10^{-2})$ eV$^{2}$. We can express the individual masses in terms of the lightest mass $m_{1}$ and the independent mass-squared differences as,
\begin{equation}
m_{2}=\sqrt{m^{2}_{1}+\msol^2}, \hspace{0.2cm}
m_{3} =\sqrt{m^{2}_{1}+\matm^2}, \hspace{0.2cm}
m_{4}=\sqrt{m^{2}_{1}+\Delta m^{2}_{s_2}}, \hspace{0.2cm}
m_{5} =\sqrt{m^{2}_{1}+\Delta m^{2}_{s_1}}. \label{eq:mass_order_SSN}
\end{equation}

In Eq.~\eqref{eq:mass_order_SSN}, we have used the absolute values of $\matm^2, \mso^2,\mst^2$ and relative sign is taken care of internally and this will also be followed for the other mass orderings.

\item Sterile-Sterile-Inverted Hierarchy Scheme (SSI): 
In this scheme, $m_{5}>m_{2}>m_{1}>m_{4}>m_{3}$ for ($\Delta m^{2}_{s_2}<\matm^2$) and $m_{5}>m_{4}>m_{2}>m_{1}>m_{3}$ for ($\Delta m^{2}_{s_2}>\matm^2$) . The mass-squared difference of the heaviest neutrino mass state and the lightest neutrino mass state is given by $\Delta m^{2}_{s_1}\equiv m^{2}_{5}-m^{2}_{3}=1.3$ eV$^{2}$. The mass-squared difference of the second sterile neutrino mass state with the lightest mass is varied as $\Delta m^{2}_{s_2}\equiv m^{2}_{4}-m^{2}_{3}=10^{-4} \text{ or }  10^{-2}$ eV$^{2}$. We can express the individual masses in terms of the lightest mass $m_{3}$ and the independent mass-squared differences as,

\begin{equation}
m_{1} =\sqrt{m^{2}_{3}+\matm^2-\msol^2},  \hspace{0.2cm}
m_{2} =\sqrt{m^{2}_{3}+\matm^2}, \hspace{0.2cm}
m_{4} =\sqrt{m^{2}_{3}+\Delta m^{2}_{s_2}},  \hspace{0.2cm}
m_{5} =\sqrt{m^{2}_{3}+\Delta m^{2}_{s_1}}.
\end{equation}

\item Sterile-Normal-Sterile Hierarchy Scheme (SNS):
In this scheme, the lightest neutrino mass state is the $m_4$ sterile state, following the mass hierarchy as $m_{5}>m_{3}>m_{2}>m_{1}>m_{4}$ . We can express the individual masses in terms of the lightest mass $m_{4}$ and the independent mass-squared differences as,
\begin{equation}
m_{1}=\sqrt{m^{2}_{4}+\Delta m^{2}_{s_2}}, \hspace{0.2cm}
m_{2} =\sqrt{m^{2}_{4}+\Delta m^{2}_{s_2}+\msol^2}, \hspace{0.2cm}
m_{3}=\sqrt{m^{2}_{4}+\Delta m^{2}_{s_2}+\matm^2}, \hspace{0.2cm}
m_{5} =\sqrt{m^{2}_{4}+\Delta m^{2}_{s_2}+\Delta m^{2}_{s_1}}.
\end{equation}

\item Sterile-Inverted-Sterile Hierarchy Scheme (SIS):
In this scheme, the lightest neutrino mass state is the $m_4$ sterile state, following the mass hierarchy as $m_{5}>m_{2}>m_{1}>m_3>m_{4}$. The individual masses are expressed in terms of the lightest mass $m_{4}$ and the independent mass-squared differences as,

\begin{equation}
m_{1} =\sqrt{m^{2}_{4}+\Delta m^{2}_{s_2}+\matm^2-\msol^2},\hspace{0.1cm}
m_{2} =\sqrt{m^{2}_{4}+\Delta m^{2}_{s_2}+\matm^2},\hspace{0.2cm}
m_{3}=\sqrt{m^{2}_{4}+\Delta m^{2}_{s_2}},\nonumber
\end{equation}
\begin{equation}
m_{5} =\sqrt{m^{2}_{4}+\Delta m^{2}_{s_1}+\Delta m^{2}_{s_2}}.
\end{equation}

Here we have considered sterile inverted scenarios only for $\mst^2 = 0.01,10^{-4}\eV^2$ and ignored the inverted orderings for $\mso^2 = 1.3~\eV^2$. This is because the inverted cases for the $\mso^2$ is already ruled by the constraints on sum of neutrino masses, effective electron mass from nuclear $\beta$ decay and effective Majorana mass from \nubb~\cite{Goswami:2024ahm}.
\end{itemize}

\section{Bounds on $3+2$ sterile neutrino scenario}\label{sec:bound_3+2}

\subsection{Sum of Neutrino Masses and Cosmological Constraints}

One of the most direct probes of the absolute neutrino mass scale arises from cosmological observations, which are sensitive to the sum of all neutrino masses,
\begin{equation}
\Sigma = \sum_{i} m_i .
\end{equation}
In the standard three-flavor case, the Planck 2018 data in combination with BAO measurements impose an upper bound of $\Sigma \lesssim 0.12$ eV (95\% C.L.), {considering the standard $\Lambda_{\text{CDM}}$ model~\cite{Planck:2018vyg}.} However, in the $3+2$ framework, we have two additional mass eigenstates $m_4$ and $m_5$ which enter the sum. {If, the active-sterile mixing is large, sterile neutrinos become fully thermalised in the early Universe, following the same Fermi-Dirac distribution as active flavors. As a result, they contribute equally to the effective number of relativistic degrees of freedom,  $N_{\text{eff}}$, and to the total mass sum which is enhanced relative to the three-flavor case,
\begin{equation}
\Sigma^{3+2} = m_1 + m_2 + m_3 + m_4 + m_5 .
\end{equation}
However, precise measurement of cosmic microwave background~(CMB) by the Planck collaboration constrain the $N_{\text{eff}}$ as~\cite{Planck:2018vyg},
\begin{eqnarray}
    N_{\text{eff}} = 2.99^{+0.34}_{-0.33}\quad~(95\% ~ \text{C.L.})
\end{eqnarray}
which ruled out the possibility of extra sterile states. Moreover, for the eV scale sterile neutrinos, $\Sigma \gtrsim 1 \,\mathrm{eV}$, far above the current cosmological upper limits.}

This tension can only be alleviated if the sterile neutrinos are not fully thermalized in the early Universe. Several mechanisms can suppress their thermalization, including low reheating temperatures, secret interactions in the sterile sector namely, new gauge or scalar interactions~\cite{Dasgupta:2013zpn, Hannestad:2013ana, Cline:2019seo, Farzan:2019yvo, Forastieri:2017oma, Yaguna:2007wi, Gelmini:2019wfp, Gelmini:2019esj, Hasegawa:2020ctq} among the sterile neutrinos themselves that redistribute their momenta and suppress free-streaming, thereby relaxing the $N_{\rm eff}$ and structure-formation constraints.

In such scenarios, the physical masses of the sterile neutrinos do not directly contribute to the sum $\Sigma$. Instead, their cosmological imprint is captured by an effective mass parameter $m_s^{\rm eff} = m_s\,\Delta N_{\rm eff}$~\cite{Hagstotz:2020ukm}, where $\Delta N_{\rm eff}$ denotes the deviation of the effective number of neutrino species from its Standard Model value, $N_{\rm eff}^{\rm SM} = 3.044 \pm 0.002$~\cite{Bennett:2020zkv, Akita:2020szl, Froustey:2020mcq}. In this case, the sum of neutrino masses can be expressed as

\begin{eqnarray}
    \Sigma^{3+2} & =& m_1+ m_2+ m_3 + \Delta N_{\text{eff}}~\left(  m_4 +m_5 \right) ,\label{eq:sum_3+2}
\end{eqnarray}
where $\Delta N_{\text{eff}} = N_{\text{eff}} - N_{\text{SM}}$.
Cosmology remains a crucial test of the $3+2$ hypothesis, as even partially populated sterile states leave imprints on large-scale structure formation, the cosmic microwave background anisotropies, and the expansion history of the Universe.

\subsection{Effective Electron Neutrino Mass in $\beta$ Decay}

Another important probe of the absolute neutrino mass scale arises from precision measurements of the endpoint spectrum in single $\beta$ decay experiments, such as those performed by KATRIN. Sterile neutrinos with active-sterile mixing $|U_{ei}|^2$ can affect the beta-decay spectrum through two distinct signatures, depending on the sterile mass $m_s$ relative to the experimental energy resolution $\Delta E \simeq 0.93$~eV. For sterile masses significantly larger than the experimental resolution, $m_s \gg \Delta E$, the sterile state produces a kink-like feature in the spectrum at an energy $Q-m_s$ below the endpoint, where $Q=18.574$~keV is the tritium beta-decay endpoint energy. The amplitude of this kink-like feature is proportional to $|U_{ei}|^2$, and such a resolved spectral distortion cannot be fully described by the effective mass parameter $m_\beta$ \cite{KATRIN:2025lph}.

On the other hand, for sterile neutrino masses comparable to or smaller than the experimental resolution, $m_s \lesssim \Delta E$, the kink-like structure is not fully resolved. Instead, the sterile contribution appears as a smeared distortion of the endpoint spectrum and is effectively incorporated into the parameter
\begin{equation}
m_\beta=\sqrt{\sum_i |U_{ei}|^2 m_i^2}\, 
\label{eq:mbeta}
\end{equation}
where the sum runs over all neutrino mass eigenstates, including the sterile ones. Since the eV-scale sterile state considered in this work satisfies $m_s \sim \mathcal{O}(1)$~eV, comparable to the KATRIN resolution, its effect lies in the transition region between these two regimes and is expected to
manifest primarily through a partially smeared endpoint distortion that must be extracted through a differential spectral fit rather than a sharply resolved kink. KATRIN has provided an exclusion contour on the $\Delta m_{41}^2- \sin^2 (2\theta_{ee})$ plane using their latest data~\cite{KATRIN:2025lph} and our considered parameters are consistent with the KATRIN's allowed region.

In the standard three-flavor scenario, $m_\beta$ depends only on the active neutrino parameters, and KATRIN currently places an upper limit on $m_\beta $ as,
\begin{equation}
m_\beta < 0.45 \,\mathrm{eV} \quad (90\% \, \mathrm{C.L.}) \,,
\end{equation}
which is already close to the sensitivity required to test the quasi-degenerate regime of active neutrinos~\cite{KATRIN:2024cdt}.

In the $3+2$ framework, the effective mass receives additional contributions from the two sterile-dominated states,
\begin{equation}
m_\beta^{3+2} = \sqrt{ \, |U_{e1}|^2 m_1^2 + |U_{e2}|^2 m_2^2 + |U_{e3}|^2 m_3^2 + |U_{e4}|^2 m_4^2 + |U_{e5}|^2 m_5^2 \, } .
\end{equation}

In Table~\ref{tab:expt_inputs}, we have listed the experimental inputs including the best fit values and the corresponding $3\sigma$ ranges of the three neutrino oscillation parameters for both NO and IO. If the sterile neutrino mixings $|U_{e4}|^2, |U_{e5}|^2$ are not extremely suppressed, their contributions can significantly enhance $m_\beta$ relative to the three-flavor case. In the $3+1$ framework, stringent constraints on active–sterile mixing arise from the non-observation of sterile neutrino oscillations in the MINOS and MINOS$^+$ experiments. To the best of our knowledge, a dedicated analysis of this data in the $3+2$ scenario is not currently available. We therefore adopt the mixing constraints from Ref.~\cite{MINOS:2017cae,Acero:2022wqg}, which performed a combined analysis of MINOS, MINOS$^+$, Daya Bay, and Bugey-3 data in the $3+1$ scenario, and apply them independently to each active-sterile mixing angle; the resulting input ranges are summarised in Table~\ref{tab:theor_input}.

In the $3+1$ framework, the short-baseline electron-neutrino disappearance probability approximately takes the two-flavor form
\begin{equation}
    P(\nu_e \to \nu_e) \simeq 1 -
    4|U_{e4}|^2\!\left(1-|U_{e4}|^2\right)
    \sin^2\!\left(\frac{\Delta m_{41}^2 L}{4E}\right),
\end{equation}
so that the bound is primarily driven by a single active-sterile mixing parameter. In the $3+2$ scenario, however, the disappearance probability depends simultaneously on both $|U_{e4}|^2$ and $|U_{e5}|^2$, together with additional interference terms involving the two independent oscillation frequencies. As a result, correlations among the sterile-sector parameters can partially reduce the net disappearance signal for fixed individual mixing
angles~\cite{Karagiorgi:2009nb,Donini:2001xy}. Consequently, regions of parameter space with comparatively larger individual mixings may remain allowed in a full $3+2$ analysis relative to the minimal $3+1$ framework.

By applying the $3+1$ upper limits on $|U_{e4}|^2$ and $\sin^2\theta_{15}$ directly, we therefore adopt benchmark ranges that are likely narrower than those obtained from a dedicated $3+2$ global fit, rendering our approach conservative. The quantitative impact on $m_\beta$ is
moderate because the sterile contribution enters as $m_\beta \propto \sqrt{|U_{ei}|^2}\,m_i$. Therefore, even an order-unity relaxation of the mixing bounds would lead only to a comparatively mild enhancement of the sterile contribution to $m_\beta$, remaining within the
uncertainty ranges shown in our numerical analysis. A dedicated global fit in the full $3+2$ framework would nevertheless be important to quantify these effects more precisely and is left for future work.

\begin{table}[t]
\centering
\begin{tabular}{|c|c|c|c|c|}
\hline
Parameters & \multicolumn{2}{c|}{Normal Ordering (NO)} & \multicolumn{2}{c|}{Inverted Ordering (IO)} \\ \cline{2-5}
 & $3\sigma$ Range & Best Fit & $3\sigma$ Range & Best Fit \\ 
\hline \hline
$\sin^{2}\theta_{12}$ & $0.289 : 0.335$ & $0.309$ & $0.289 : 0.335$ & $0.309$ \\ 
\hline
$\sin^{2}\theta_{13}$ & $0.0202 : 0.0239$ & $0.0220$ & $0.0202 : 0.0239$ & $0.0220$ \\ 
\hline
$\msol^2$/$10^{-5}$ eV$^{2}$ & $7.14 : 7.86$ & $7.50$ & $7.14 : 7.86$ & $7.50$ \\ 
\hline
$\matm^2$/$10^{-3}$ eV$^{2}$ & $2.428 : 2.597$ & $2.511$ & $-2.581 : -2.408 $& $-2.498$ \\ 
\hline
\end{tabular}
\caption{3$\sigma$ ranges and best-fit values of the three neutrino oscillation parameters~\cite{Esteban:2024eli}. Here, $\msol^2 \equiv m^{2}_{2}-m^{2}_{1}$ and $\matm^2\equiv m^{2}_{3}-m^{2}_{1}$ for NO and $m^{2}_{2}-m^{2}_{3}$ for IO. For $\sin^2\theta_{12}$ we have used result from Ref.~\cite{JUNO:2025gmd}.}
\label{tab:expt_inputs}
\end{table}

\begin{table}[t]
\centering
\begin{tabular}{|c|c|c|}
\hline
Parameters & Case 1 & Case 2 \\ 
\hline \hline
$\Delta m_{s_1}^{2}$ & 1.3 eV$^{2}$  & 1.3 eV$^{2}$  \\ 
$\sin^{2}\theta_{15}$ & $0.001 : 0.01$ & $0.001 : 0.01$ \\ 
\hline
$\Delta m_{s_2}^{2}$ & $10^{-2}$ eV$^{2}$  & $10^{-4}$ eV$^{2}$   \\ 
$\sin^{2}\theta_{14}$ &  $5 \times 10^{-4}$ :  $5 \times 10^{-3}$ & $0.1 : 0.2$  \\ 
\hline
\end{tabular}
\caption{Input parameter ranges for the $3+2$ model used in this analysis. The $\sin^2\theta_{ij}$ ($i=1,\,j=4,5$) values are chosen to be consistent with the upper limits obtained from the combined MINOS, MINOS$^+$, Daya Bay, and Bugey-3 analyses of Refs.~\cite{MINOS:2017cae,Acero:2022wqg} performed in the $3+1$ framework. As discussed in the text, applying these $3+1$ bounds independently to each sterile mixing angle provides a conservative benchmark. In the full $3+2$ scenario, the disappearance probability depends on both $|U_{e4}|^2$ and $|U_{e5}|^2$ together with interference effects involving the two oscillation frequencies, allowing parameter correlations that can relax the individual mixing constraints relative to the minimal $3+1$ case.} 
\label{tab:theor_input}
\end{table}

\subsection{Effective Majorana Mass in Neutrinoless Double $\beta$ Decay}\label{ssec:0nbb-gen}

A unique probe of the Majorana nature of neutrinos and of the lepton number violation is $0\nu\beta\beta$. If observed, this process would demonstrate that neutrinos are Majorana fermions and provide insights into the absolute neutrino mass scale and the underlying new physics mechanisms. The \nubb ~ is governed by $m_{\beta\beta}$ which can be expressed for the $3+2$ scenario as, 
\begin{equation}
m_{\beta\beta} =  \sum_{i=1}^{5} U_{ei}^2 \, m_i , \label{eq:mbb-gen}
\end{equation}
where $m_i$ are the neutrino mass eigenvalues, $U_{ei}$ are the elements of the first row of the PMNS mixing matrix, and $\alpha_i$ denote the Majorana CP phases. We can also express Eq.~\eqref{eq:mbb-gen} in terms of neutrino masses and mixing angles as,
\begin{eqnarray}
    m_{\beta\beta} &= & \, c_{12}^2 c_{13}^2 c_{14}^2 c_{15}^2 m_1 + s_{12}^2 c_{13}^2 c_{14}^2 c_{15}^2 m_2 e^{i\alpha_{21}} + s_{13}^2 c_{14}^2 c_{15}^2 m_3 e^{i\alpha_{31}} + s_{14}^2 c_{15}^2 m_4 e^{i\alpha_{41}} + s_{15}^2 m_5 e^{i\alpha_{51}} \, \label{eq:mbb-3+2} .
\end{eqnarray}
Here, $\alpha_{21}$, $\alpha_{31}$, $\alpha_{41}$, and $\alpha_{51}$ are independent Majorana phases. The additional terms proportional to $m_4$ and $m_5$ can substantially modify the phenomenology of $0\nu\beta\beta$. {In particular, if the sterile neutrinos have masses of the order of eV (sub-eV) scale and mix appreciably with the electron flavor, their contributions can enhance or suppress $m_{\beta\beta}$ depending on the relative CP phases.} This implies that the standard lower bounds on $m_{\beta\beta}$ for inverted ordering may no longer hold, and cancellations can occur in regions of parameter space that would otherwise predict a signal within the reach of upcoming experiments.

Most stringent upper limit on $m_{\beta\beta}$ is placed by the KamLAND-Zen~\cite{KamLAND-Zen:2024eml} experiments as,
\begin{equation}
m_{\beta\beta} \lesssim (0.028 - 0.122) \,\mathrm{eV}. 
\end{equation}
The range in the $m_{\beta\beta}$ values are subject to the nuclear matrix element uncertainties. In the presence of sterile states, these bounds apply directly to $m_{\beta\beta}^{3+2}$, thereby constraining the active-sterile mixing parameters $U_{e4}, U_{e5}$ and the sterile neutrino masses $m_4, m_5$.

An important aspect of the $3+2$ scenario is that the additional sterile contributions can significantly modify the structure of the effective Majorana mass parameter relative to the standard three-neutrino framework. Depending on the sterile masses, mixing angles, and CP phases, the active and sterile contributions to $m_{\beta\beta}$ may interfere either constructively or destructively, thereby altering the allowed parameter regions associated with different neutrino mass orderings. Consequently, the interpretation of
neutrinoless double $\beta$ decay constraints can differ substantially from the minimal three-flavor case. Future ton-scale experiments such as LEGEND-1000, with sensitivity reaching the $\mathcal{O}(10^{-3})\,\mathrm{eV}$ scale, will therefore provide complementary probes of light sterile-neutrino scenarios. The quantitative implications for the various mass-ordering schemes considered in this work are discussed in Sec.~\ref{sec:results}.

\section{Results and Discussion}\label{sec:results}
For our numerical analysis, we have used the input parameters from Tables~\ref{tab:expt_inputs}, \ref{tab:theor_input}, and  \ref{tab:ue_values}. Entries in Table~\ref{tab:ue_values} denote the $3\sigma$ ranges of the mixing matrix elements $|U_{ei}|^2$ ($i=1$–$5$) in the $3+2$ sterile neutrino framework. The active components $|U_{e1}|^2$, $|U_{e2}|^2$, and $|U_{e3}|^2$ are obtained from global oscillation fits, while the sterile admixtures $|U_{e4}|^2$ and $|U_{e5}|^2$ correspond to benchmark values consistent with MINOS, MINOS$^{+}$, Daya Bay, and Bugey-3 constraints for the indicated sterile mass-squared differences.

\begin{table}[t]
\centering
\resizebox{\columnwidth}{!}{
\begin{tabular}{|c|c|c|c|c|c|c|c|c|}
\hline
\multicolumn{2}{|c|}{$|U_{e1}|^{2} = c_{12}^2c_{13}^2c_{14}^2c_{15}^2 $} & \multicolumn{2}{c|}{$|U_{e2}|^{2} = s_{12}^2c_{13}^2c_{14}^2c_{15}^2$} & \multicolumn{2}{c|}{$|U_{e3}|^{2} = s_{13}^2c_{14}^2c_{15}^2$} & \multicolumn{2}{c|}{$|U_{e4}|^{2} = s_{14}^2c_{15}^2$} & $|U_{e5}|^{2} = s_{15}^2$ \\ \cline{1-8}
$\Delta m_{s_2}^2=10^{-4}$ & $\Delta m_{s_2}^2=10^{-2}$  & $\Delta m_{s_2}^2=10^{-4}$ & $\Delta m_{s_2}^2=10^{-2}$ & $\Delta m_{s_2}^2=10^{-4}$ & $\Delta m_{s_2}^2=10^{-2}$ & $\Delta m_{s_2}^2=10^{-4}$ & $\Delta m_{s_2}^2=10^{-2}$ & $\Delta m_{s_1}^2=1.3$ eV$^{2}$ \\ 
\hline \hline
$0.52 : 0.63$& $0.64 : 0.701$ & $0.22 : 0.29$& $0.27 :0.33$ & $0.016 : 0.021$ & $0.02 : 0.024$ & $0.099 : 0.1997$ & $\left( 0.495 :4.99\right) \times 10^{-3}$ & $0.001 : 0.01$ \\ 
\hline
\end{tabular}}
\caption{Representative $3\sigma$ ranges of the mixing matrix elements $|U_{ei}|^2$ ($i=1$–$5$) in the $3+2$ sterile neutrino framework.}
\label{tab:ue_values}
\end{table}


\subsection{Cosmological Constraints on the $3+2$ Scenario}

Cosmological observations provide some of the most stringent bounds on light sterile neutrinos through their impact on the sum of neutrino masses $\Sigma$ and on the effective number of relativistic species $N_{\rm eff}$. In this analysis, we have considered the constraint obtained from the combined analysis of Planck + BAO + Hubble parameter measurement + Supernova Ia data, fitted with a ten-parameter cosmological model (10-PCM)~\cite{Acero:2022wqg} i.e.
\[
\Lambda_{\rm CDM} + N_{\rm eff} + m_s^{\rm eff} + \omega_0 + n_{\rm run},
\] as, 
\begin{eqnarray}
    N_{\rm eff} = 3.11^{+0.37}_{-0.36}, \qquad  \Sigma = 0.16~\text{eV}. \label{eq:10-pcm-limit}
\end{eqnarray} Here $\omega_0$ is the equation of state parameter of dark energy, and $n_{\rm run}$ denotes the running of the scalar spectral index which is related to the initial conditions of the Universe.

\begin{figure}[t]
\centering
\includegraphics[scale=0.45]{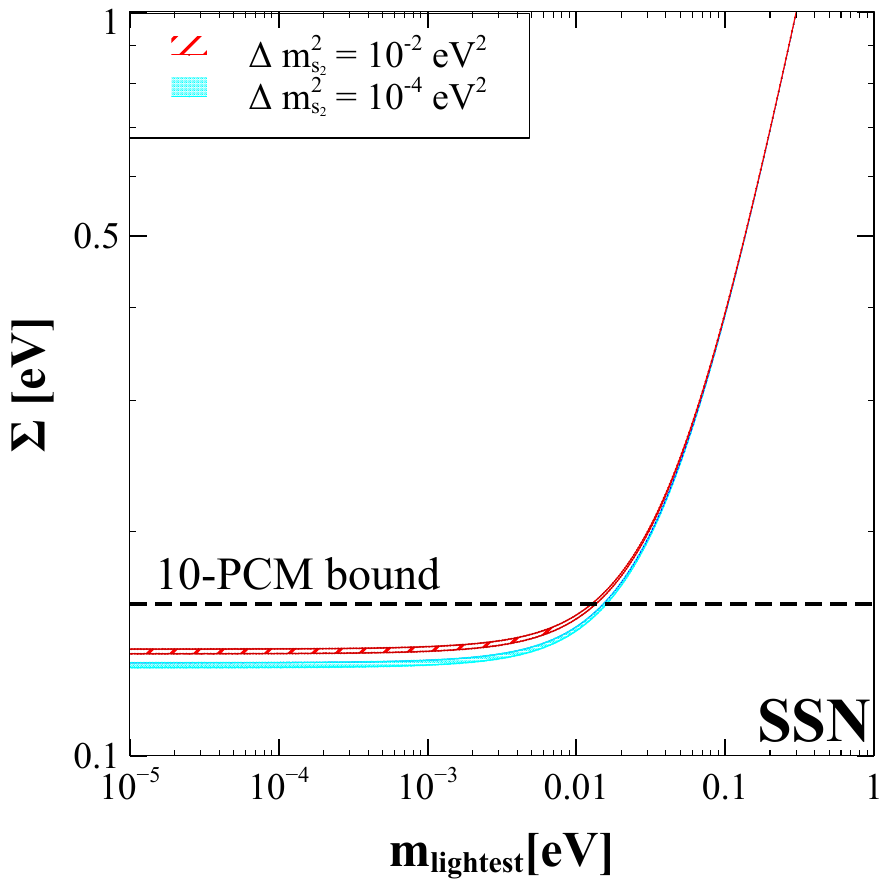}
\includegraphics[scale=0.45]{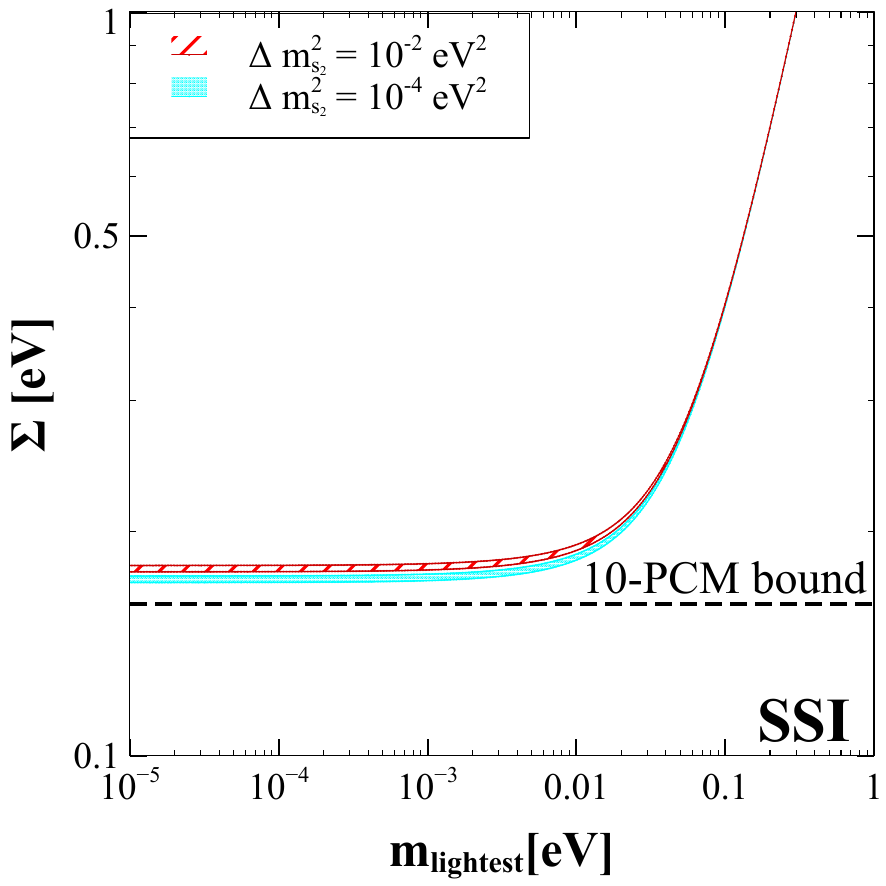}
\includegraphics[scale=0.45]{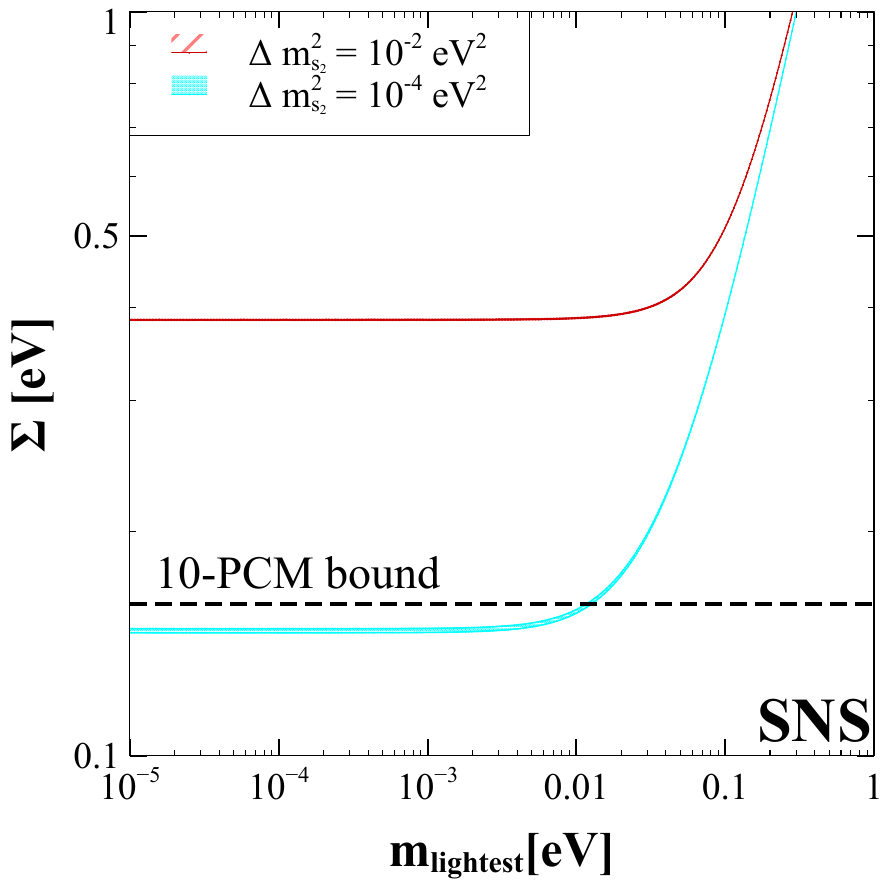}
\includegraphics[scale=0.45]{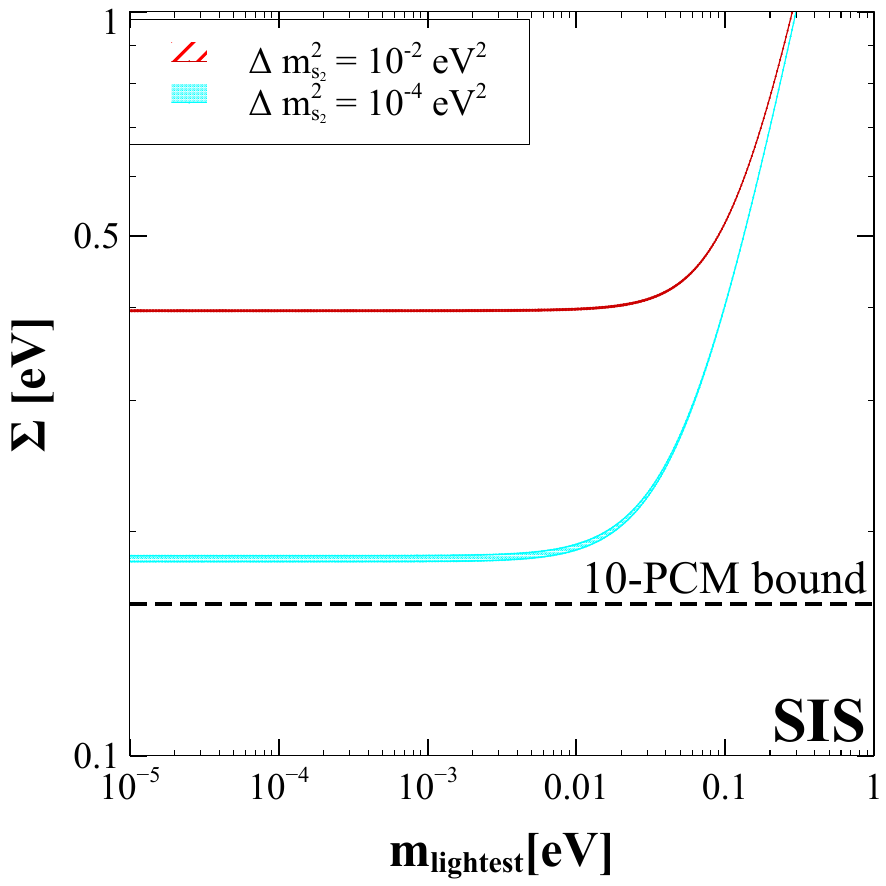}
\caption{Total effective mass $\Sigma$ vs the lightest neutrino mass $m_{\rm lightest}$ in different scenarios: SSN (top-left), SSI (top-right), SNS (bottom-left), and SIS (bottom-right). The red-slashed and cyan colors correspond to $\Delta m_{s_2}^2 = 10^{-2}\, \&\, 10^{-4}$ $\rm eV^2$, respectively. The horizontal black dashed line corresponds to the bound~(Eq.~\eqref{eq:10-pcm-limit}) from the 10 parameter cosmological (10-PCM) model.}
\label{fig:sigma_cosmo}
\end{figure}

In Fig.~\ref{fig:sigma_cosmo}, we present $\Sigma^{3+2}$ as a function of the lightest neutrino mass for the different benchmark mass-splitting schemes considered. The figure illustrates that the presence of extra sterile states in some scenarios substantially enhances the sum of neutrino masses compared to the three-flavor case, highlighting the tension with the cosmology in the $3+2$ scenario. Our observations in Fig.~\ref{fig:sigma_cosmo} are as follows:
\begin{itemize}
    \item The SSN scenario is allowed by the 10-PCM bound up to $m_{\rm lightest} \sim 0.016~(0.013)$ eV for $\Delta m_{s_2}^2 = 10^{-4}~(10^{-2})~\rm eV^2$. On the other hand, the SSI scenario is disfavored by the 10-PCM bound for both mass-squared differences. This behavior is expected because in the SSI scenario, for vanishing lightest neutrino mass, one has $m_1 \approx m_2 \approx 0.05~\eV$. In addition, the effective sterile contribution to the cosmological mass bound, $\Delta N_{\rm eff}\, m_5$~[c.f Eq.~\eqref{eq:sum_3+2}], is approximately $0.08~\eV$, which already exceeds the 10-PCM limit even before including the contribution from $m_4$.
    
    \item The SNS scenario is disfavored for $\Delta m_{s_2}^2 = 10^{-2}$ $\rm eV^2$ however, it is allowed for $\Delta m_{s_2}^2 = 10^{-4}$ $\rm eV^2$ up to $m_{\rm lightest} \sim 0.012$ eV as far as the 10-PCM bound is concerned. On the other hand, the SIS scenario is completely disfavored by the 10-PCM bound like the SSI scenario. The above results are summarised in Table~\ref{tab:10pcm}.
\end{itemize}

\begin{table}
\centering
\begin{tabular}{|c|c|c|}
\hline
Mass ordering & \multicolumn{2}{c|}{$\Delta m^{2}_{s_1}=1.3$ eV$^{2}$}  \\ \cline{2-3}
& $\Delta m^{2}_{s_2}=10^{-4}$ eV$^{2}$ & $\Delta m^{2}_{s_2}=10^{-2}$ eV$^{2}$ \\ 
\hline \hline
SSN  & $<0.016$ & $<0.013$  \\ 
\hline
SSI  & Disallowed & Disallowed \\ 
\hline
SNS & $<0.012$ & Disallowed  \\ 
\hline
SIS & Disallowed & Disallowed  \\ 
\hline
\end{tabular}
\caption{Summary of four mass spectra for two different combinations of $\Delta m^{2}_{s_1}$ and $\Delta m^{2}_{s_2}$ within the limit set by the 10-PCM model in cosmology. The limit corresponds to the value of $m_{\rm lightest}$ up to which the scenario is allowed.}
\label{tab:10pcm}
\end{table}

\subsection{Implications for $\beta$ Decay in the $3+2$ Scenario}

In the $3+2$ framework, $m_\beta$ receives additional contributions from the two sterile states, in addition to the standard three active neutrinos. In our numerical analysis, we have varied all oscillation parameters within their respective $3\sigma$ allowed intervals as reported in Table~\ref{tab:ue_values}, together with the benchmark sterile splittings $\Delta m^2_{s_1} = 1.3~\text{eV}^2$ and $\Delta m^2_{s_2} = 10^{-2},\,10^{-4}~\text{eV}^2$, as well as the associated active--sterile mixing angles.

\begin{figure}[t]
\centering
\includegraphics[scale=0.45]{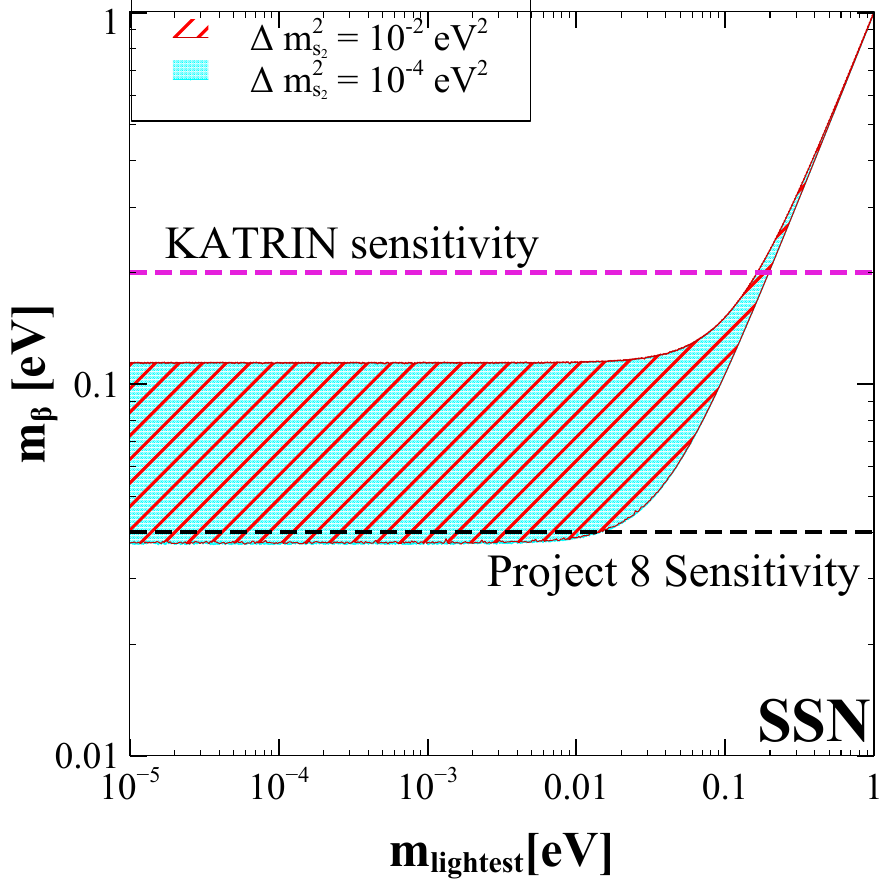}
\includegraphics[scale=0.45]{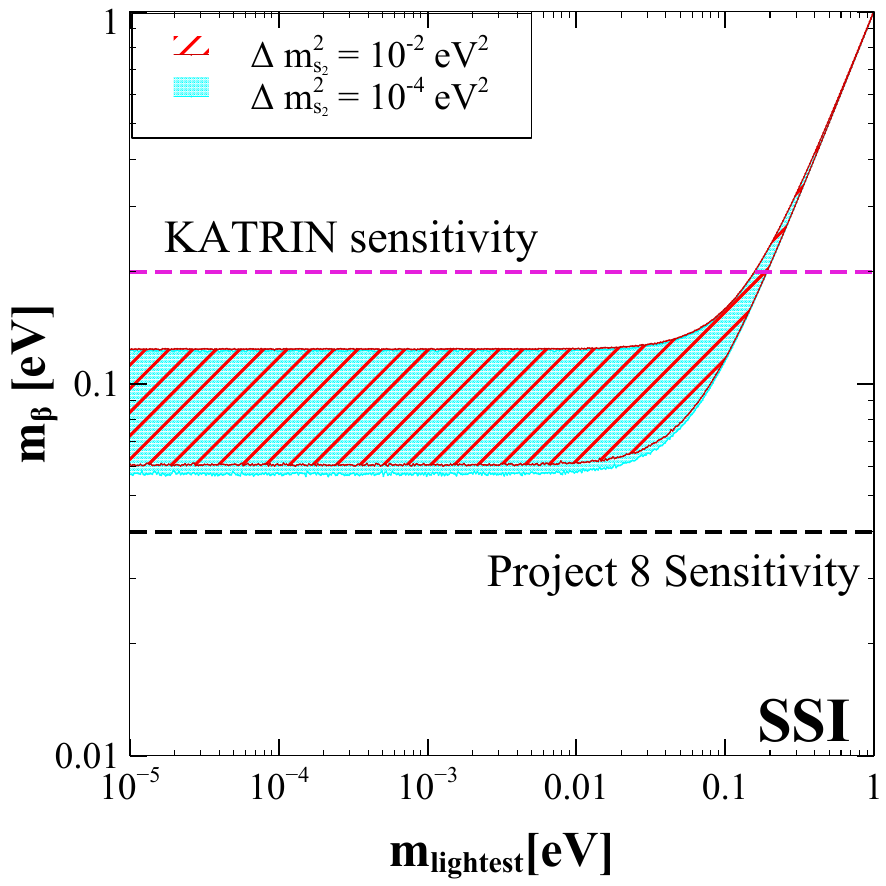}
\includegraphics[scale=0.45]{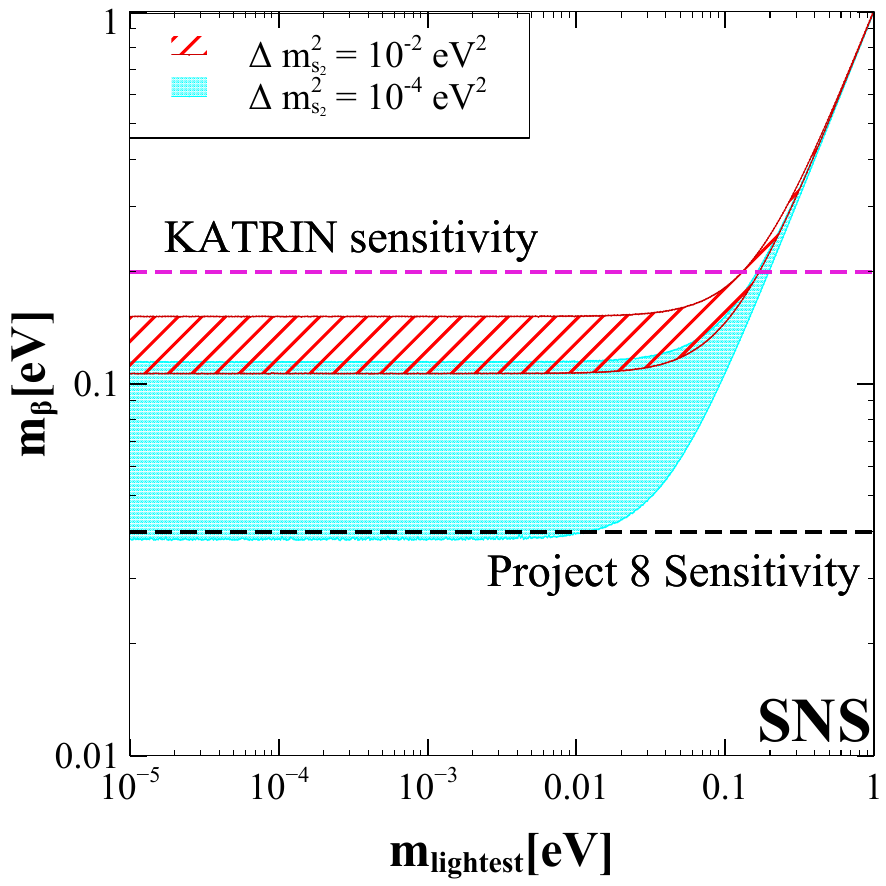}
\includegraphics[scale=0.45]{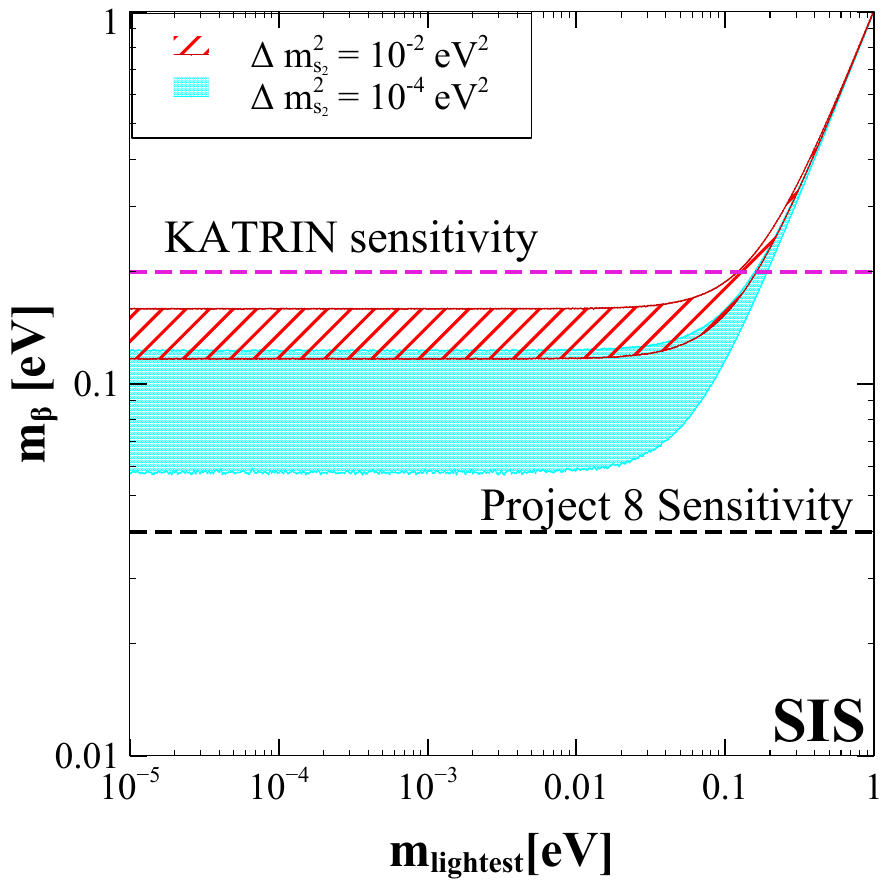}
\caption{The figure shows the variation of $m_{\beta}$ as a function of the lightest neutrino mass for the $3+2$ spectrum. The red-slashed and cyan bands correspond to $\Delta m_{s_2}^2 = 10^{-2} \text{ and } 10^{-4}~\eV^2$ mass square differences. The magenta~(black) dashed line shows the projected sensitivity of KATRIN~(Project~8) experiment.}
\label{fig:mbeta}
\end{figure}

Fig.~\ref{fig:mbeta} shows the resulting predictions for $m_\beta$ as a function of the lightest neutrino mass for the mass-ordering schemes considered in this study. Since the eV-scale sterile state considered here lies close to the KATRIN energy resolution, the treatment in terms of the  effective mass parameter $m_\beta$ should be regarded as conservative, as a dedicated kink analysis could in principle provide additional sensitivity to the sterile contribution. The projected sensitivity from KATRIN is taken to be $0.2$~eV, while the anticipated reach of the Project~8 experiment is $m_\beta \sim 40$~meV~\cite{Project8:2022wqh}.

We observe that the presence of two sterile states can shift the predicted $m_\beta$ spectrum upwards compared to the standard three-neutrino and $3+1$ cases. Our main observations are as follows:

\begin{itemize}
    \item In the SSN scenario, the behavior of $m_\beta$ remains similar over the entire range of $m_{\text{lightest}}$. This is because the eV-scale sterile neutrino contribution dominates for both mass-squared differences. Consequently, $m_\beta$ can be approximated as, $m_\beta \approx \sqrt{U_{e5}^2m_5^2}$ yielding values in the range $\sim 0.03~\text{eV}$ to $0.13~\text{eV}$, as illustrated in the upper-left panel of Fig.~\ref{fig:mbeta}.

    \item In the SSI scenario, the maximum value of $m_\beta$ is also dominated by the eV-scale sterile contribution. However, for the minimum value, the contribution from the eV-scale sterile state becomes comparable to that of the two active neutrino states ($m_1$ and $m_2$). As a result, the minimum value of $m_\beta$ is slightly shifted toward higher values compared to the SSN scenario.

    \item For the SNS and SIS scenarios, the behavior of $m_\beta$ differs for the two choices of $\mst^2$. In these cases, the relative ordering of the lightest neutrino mass state ($m_4$) and the active neutrino state ($m_1$) depends on the value of $\mst^2$. Consequently, $m_5$, which is defined relative to $m_1$, occupies significantly different mass scales for $\mst^2 = 10^{-2}~\text{eV}^2$ and $10^{-4}~\text{eV}^2$. This leads to larger values of $m_\beta$ for $\mst^2 = 10^{-2}~\text{eV}^2$ compared to $10^{-4}~\text{eV}^2$. Additionally, the similarity in the $m_\beta$ behavior between SSN and SNS, and between SSI and SIS, indicates that the dominance of the eV-scale sterile state is such that the normal or inverted ordering of the other sterile state becomes irrelevant in these scenarios.

    \item Although the current bound ($m_\beta < 0.45~\text{eV}$) and the projected KATRIN sensitivity ($m_\beta < 0.2~\text{eV}$) are compatible with all the mass-ordering schemes up to $m_{\text{lightest}}\simeq 0.15~\eV$, several of these scenarios are expected to be probed significantly by the projected sensitivity of the Project~8 experiment. In particular, the anticipated reach of Project~8 in $m_\beta$ would allow substantial regions of the corresponding parameter space to be tested.

    \item From Fig.~\ref{fig:mbeta}, the importance of the Project~8 experiment becomes evident. The projected sensitivity of Project~8 is sufficient to exclude the SSI and SIS scenarios entirely. Furthermore, it can exclude the SNS scenario for $\mst^2 = 10^{-2}~\text{eV}^2$, while probing a substantial portion of the parameter space of the SSN and SNS scenarios for $\mst^2 = 10^{-4}~\text{eV}^2$. 
\end{itemize}

\subsection{Implications for Neutrinoless Double $\beta$ Decay in the $3+2$ Scenario}

Neutrinoless double $\beta$ decay not only probes the Majorana nature of the neutrinos but is also sensitive to the absolute neutrino mass scale unlike the cosmological observations and tritium $\beta$ decay, which are only sensitive to the absolute neutrino mass scale. As discussed in section~\ref{ssec:0nbb-gen}, \nubb ~ put a constraint on the half-life of the decaying isotope. This eventually provides information about the effective mass $m_{\beta \beta}$. Recall that the explicit expression for effective Majorana mass governing \nubb ~ for three flavor scenario is given by~[putting $\theta_{14}=\theta_{15} = 0$ in Eq.~\eqref{eq:mbb-3+2}],

\begin{equation}
m_{\beta\beta}^{\rm std} = \left| m_1 c_{12}^2 c_{13}^2 + m_2 s_{12}^2 c_{13}^2 e^{i\alpha_{21}} + m_3 s_{13}^2 e^{i\alpha_{31}} \right|, \label{eq:mbb_standard}   
\end{equation}

where $\alpha_{21}$ and $\alpha_{31}$ are the Majorana phases. Its behavior crucially depends on the neutrino mass ordering. In the case of NO ($m_1 < m_2 < m_3$), for $m_{\rm lightest} \ll \sqrt{\Delta m^2_{\rm sol}}, \sqrt{\Delta m^2_{\rm atm}}$, the masses satisfy $m_2\simeq \sqrt{\Delta m^2_{\rm sol}}\simeq 0.01~\mathrm{eV}$ and $m_3\simeq \sqrt{\Delta m^2_{\rm atm}}\simeq 0.05~\mathrm{eV}$. The effective mass can then be approximated as $m_{\beta\beta}^{\rm NO}\simeq m_3 c_{13}^2\left|\sqrt{r}\, s_{12}^2 e^{i\alpha_{21}} + t_{13}^2 e^{i\alpha_{31}}\right|$, where $r=\Delta m^2_{\rm sol}/\Delta m^2_{\rm atm}$. A complete cancellation between the two terms would require $\sqrt{r}s_{12}^2=t_{13}^2$, which is not realized for current best-fit parameters ($\sqrt{r}s_{12}^2\simeq 0.05$, $t_{13}^2\simeq 0.02$). Consequently, $m_{\beta\beta}^{\rm std-NO}$ typically lies in the range $\left( 1\!-\!4\right) \times10^{-3}$ eV. Nevertheless, a narrow cancellation region exists around $m_{\rm lightest}\sim (2\!-\!7) \times10^{-3}$ eV, where destructive interference among the three contributions can suppresses the effective mass down to $\sim10^{-4}$ eV for specific choices of the Majorana phases ($\alpha_{21}=\alpha_{31}=\pi$). In contrast, for IO ($m_3 < m_1 < m_2$), with $m_3\approx0$ and $m_1\simeq m_2\simeq \sqrt{\Delta m^2_{\rm atm}}\simeq 0.05$ eV,
the effective mass reduces to $m_{\beta\beta}^{\rm std-IO}\simeq \sqrt{\Delta m^2_{\rm atm}}\, c_{13}^2|c_{12}^2 + s_{12}^2 e^{i\alpha_{21}}|$. Since the two dominant terms cannot fully cancel, $m_{\beta\beta}^{\rm IO}$ is bounded within $0.02~\mathrm{eV}\lesssim m_{\beta\beta}^{\rm IO}\lesssim 0.05~\mathrm{eV}$, depending on the Majorana phases $\alpha_{21,31}$. 

In the following sections, we discuss the implications for \nubb ~in the extended $3+2$ model under the four distinct scenarios: SSN, SSI, SNS, and SIS.

\subsubsection{SSN}
In the $(3+2)$ framework, the effective Majorana mass governing \nubb ~ can be expressed as,
\begin{equation}
m_{\beta\beta}^{\rm SSN} 
 =  c_{14}^2\,c_{15}^2\,\left| 
m_{\beta\beta}^{\rm Std-NO}
+ t_{14}^2\,m_4\,e^{i\alpha_{41}}
+ t_{15}^2\,m_5\,e^{i\alpha_{51}} 
\right|,
\label{eq:mbb-ssn}
\end{equation}
where $m_{\beta\beta}^{\rm Std-NO}$ represents the effective Majorana mass in the standard three-flavor framework with NO,
and the last two terms denote additional contributions from the sterile mass eigenstates $m_4$ and $m_5$. Here $t_{ij} \equiv \tan\theta_{ij}$ and  $\alpha_{41}$, $\alpha_{51}$ are the additional Majorana-type CP phases associated with the sterile neutrinos. The overall factor $c_{14}^2 c_{15}^2$ arises from the extended PMNS matrix normalization. 

\begin{figure}[t]
\centering
\includegraphics[scale=0.5]{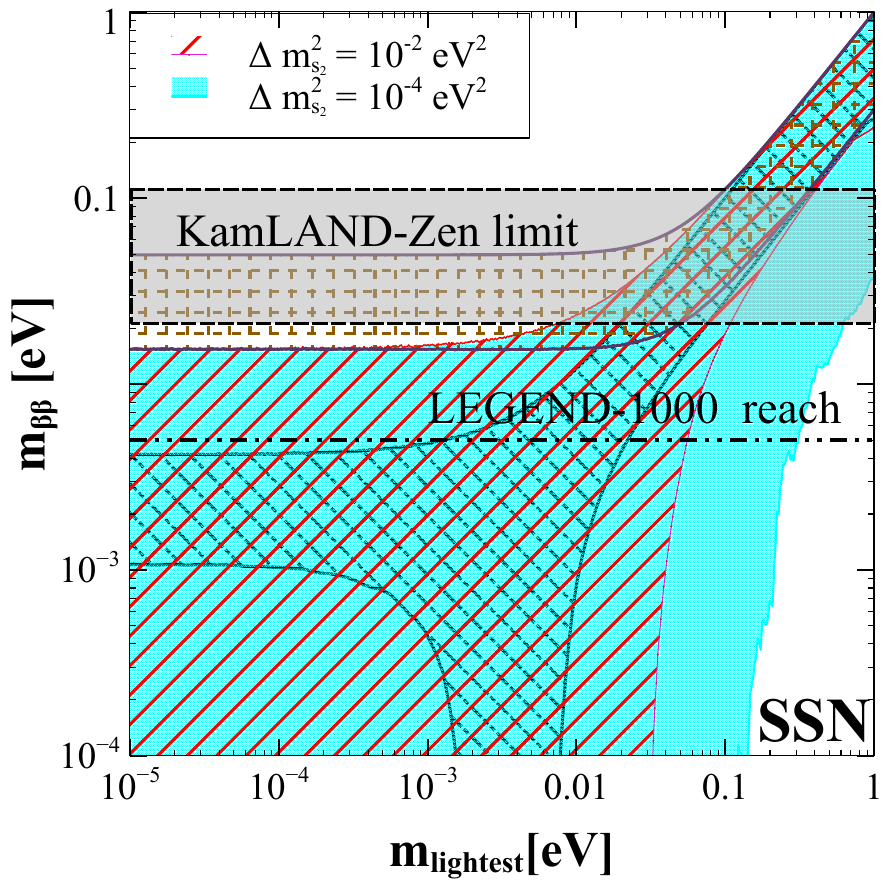}
\includegraphics[scale=0.5]{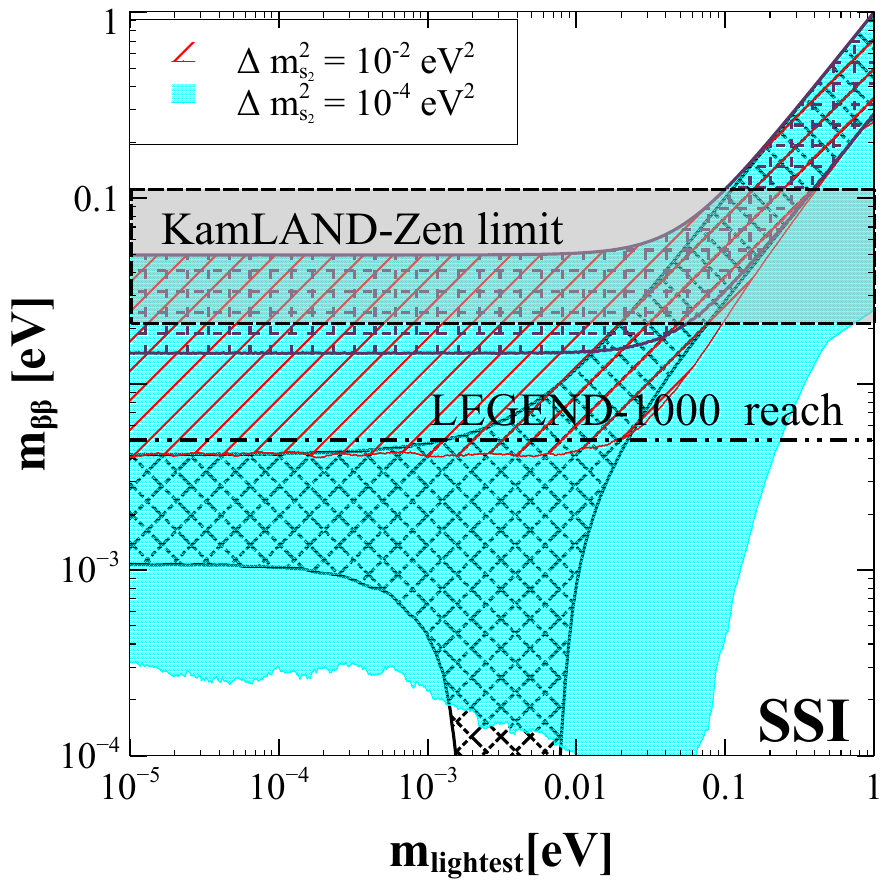}
\caption{Variation of $m_{\beta\beta}$ as a function of the light neutrino mass is shown for SSN (left panel) and SSI (right panel) scenarios. The black diagonal and vertical cross-filled regions denote variation in the standard NO and IO cases. The cyan and red-slashed band represents the values for $\mst^2 = 10^{-4}~\eV^2$ and $\mst^2 = 10^{-2}\eV^2$. Additionally, the grey band represents the current KamLAND-Zen exclusion region and the black dashed-dot line shows the future LEGEND-1000 sensitivity. }
\label{fig:SSN-SSI}
\end{figure}

The individual contributions to Eq.~(\ref{eq:mbb-ssn}) for various representative values of the lightest neutrino mass $m_{\rm lightest} = m_1$ are summarised in Table~\ref{tab:mbb-ssn}. These ranges are computed by varying the oscillation parameters over their $3\sigma$ limits (see Tabs.~\ref{tab:expt_inputs}-~\ref{tab:ue_values}). The maximum (minimum) values of $m_{\beta\beta}^{\rm SSN}$ correspond to the constructive (destructive) interference among the three terms in Eq.~(\ref{eq:mbb-ssn}), which occur for $\alpha_{51}=\alpha_{41}=0$ ($\alpha_{51}=\alpha_{41}=\pi$). The variation of $m_{\beta\beta}$ for SSN scenario is shown in the left panel of Fig.~\ref{fig:SSN-SSI}. The qualitative behavior across various regions of $m_1$ can be summarized as follows:

\begin{itemize}

\item  When $m_1 \approx 0$,   
the active-sector contribution $m_{\beta\beta}^{\rm NO}$ is typically at the level of $(1.2-4.1)\times10^{-3}$ eV, whereas the eV-scale sterile term $|t_{15}^2 m_5|$ lies in the range $(1.14-11.52)\times10^{-3}$ eV and can dominate over the other sterile contribution.  
The lighter sterile term $|t_{14}^2 m_4|$ is sub leading, especially when $\Delta m^2_{s_2}=10^{-2}$ eV$^2$, where it reaches only $\mathcal{O}(10^{-5})$ eV and for $\mst^2 = 10^{-4}\eV^2$, it varies as $(1-2) \times 10^{-3}$.  
As a result, complete cancellation of $m_{\beta\beta}^{\rm SSN}$ can occur when the relative phase $\alpha_{51}$ is close to $\pi$ and the active–sterile mixing $t_{15}^2$ is chosen near its lower bound. In this case, the maximum value of $m_{\beta\beta}^{\rm SSN}$ can go up to $0.017$ eV for $\alpha_{41,51} = 0$.

\item For the intermediate region ($m_1 \approx \sqrt{\Delta m^2_{\rm sol}}$),  
the active contribution can be as high as $\mathcal{O}(10^{-2})$ eV, becoming comparable to the eV scale sterile-induced terms. For $\Delta m^2_{s_2}=10^{-2}$ eV$^2$, the $m_4$ contribution ($|t_{14}^2 m_4|\sim 10^{-5}- 10^{-4}$ eV) remains subdominant, and the interference pattern is largely dictated by the interplay between $m_{\beta\beta}^{\rm Std-NO}$ and $\left|t_{15}^2m_5\right|$.  
In contrast, for $\Delta m^2_{s_2}=10^{-4}$ eV$^2$ where $|t_{14}^2 m_4|$ can reach $(1.3-2.7)\times10^{-3}$ eV, the $|t_{14}^2 m_4|$ term becomes comparable to the active contribution.  
In this case, we can have substantial destructive interference among the three terms in Eq.~\eqref{eq:mbb-ssn}, allowing $m_{\beta\beta}^{\rm SSN}$ to be suppressed down to a few $10^{-4}$ eV.

\item For quasi-degenerate region ($m_1 \gtrsim 0.1$ eV),  
the active component dominates the sum, yielding $m_{\beta\beta}^{\rm Std-NO}\simeq \left(2.97-10.04\right) \times 10^{-2}$ eV. The sterile terms, contributing at the $\sim \left( 10^{-3}- 10^{-2} \right)$ eV level, act as small perturbations to the active amplitude. However, in case of $\mst^2 = 10^{-4}\, \eV^2$, the cancellation can be possible between $m_{\beta\beta}^{\rm Std-NO}$ and the sterile contributions for maximum values of sterile contribution and minimum values of $m_{\beta\beta}^{\rm Std-NO}$ even around $0.1~\eV$. The required phase alignment in this case will be $\alpha_{41}\simeq \alpha_{51}\simeq \pi$.  
Beyond $m_{\rm lightest }\sim 0.2 ~\eV$, the overall $m_{\beta\beta}^{\rm SSN}$ is dominated by active neutrino contribution, with mild distortions depending on sterile contributions.
\end{itemize}

\begin{table}[t]
\centering
\begin{tabular}{|c|c|c|c|c|}
\hline
Region & $m_{\beta\beta}^{\rm Std-NO}$ (eV) & \multicolumn{2}{c|}{$|t_{14}^2 m_4|$ (eV)}  & $|t_{15}^2 m_5|$ (eV) \\
\cline{3-4} 
 &  & ($\Delta m^2_{s_2}=10^{-4}\eV^2$) & ($\Delta m^2_{s_2}=10^{-2}\eV^2$) & ($\Delta m^2_{s_1}=1.3~\eV^2$)\\ 
 \hline \hline
$m_1 \approx 0$ & $ \left(1.2 : 4.1 \right)\times 10^{-3}$ 
  & $\left(1.00:2.02\right)\times10^{-3}$ 
  & $\left(0.5:5.05\right)\times10^{-4}$ 
  & $\left(1.14:11.52\right)\times10^{-3}$ \\
\hline
$m_1 \approx \sqrt{\Delta m^2_{\rm sol}}$ 
 & $\left( 0.45 : 10.86 \right) \times 10^{-3} $
 & $\left(1.324 :2.672\right)\times10^{-3}$ 
 & $\left(0.5:5.07\right)\times10^{-4}$ 
 & $\left(1.14:11.52\right)\times10^{-3}$ \\
\hline
$m_1 \approx 0.1\ \mathrm{eV}$ & $ \left( 2.97 : 10.04 \right) \times 10^{-2}$ 
 & $\left(1.01:2.03\right)\times10^{-2}$ 
 & $\left(0.71:7.14\right)\times10^{-4}$ 
 & $\left(0.115:1.156\right)\times10^{-2}$ \\
\hline
\end{tabular}
\caption{Representative numerical ranges for $m_{\beta\beta}^{\text{SSN}}$ in the active-sector and sterile contributions in the 3+2 SSN scenario.}
\label{tab:mbb-ssn}
\end{table}

\subsubsection{SSI}
In the SSI scenario, the lightest active mass $m_3$ lies below the active and sterile states. The effective Majorana mass governing \nubb ~ can be expressed as
\begin{equation}
m_{\beta\beta}^{\rm SSI} 
 =  c_{14}^2\,c_{15}^2\,\left| 
m_{\beta\beta}^{\rm Std-IO}
+ t_{14}^2\,m_4\,e^{i\alpha_{41}}
+ t_{15}^2\,m_5\,e^{i\alpha_{51}} 
\right|,
\label{eq:mbb-ssi}
\end{equation}
where $m_{\beta\beta}^{\rm Std-IO}$ represents the effective Majorana mass in the standard three-flavor framework with IO, and the last two terms denote additional sterile mass eigenstates $m_4$ and $m_5$. Upon considering the active-sterile mixing parameters from Tabs.~\ref{tab:expt_inputs}--\ref{tab:ue_values}, the individual contributions to Eq.~\eqref{eq:mbb-ssi} for various representative values of the lightest neutrino mass $m_{\rm lightest} = m_3$ are summarized in Table~\ref{tab:mbb-ssi}. As mentioned, these ranges are computed by varying the oscillation parameters over their $3\sigma$ limits and the maximum (minimum) values of $m_{\beta\beta}^{\rm SSI}$ correspond to the constructive (destructive) interference among the three terms in Eq.~(\ref{eq:mbb-ssi}), which occur for $\alpha_{51}=\alpha_{41}=0$ ($\alpha_{51}=\alpha_{41}=\pi$).
The right panel of Fig.~\ref{fig:SSN-SSI} shown the behavior of $m_{\beta\beta}^{\text{SSI}}$ as a function of the lightest neutrino state. The qualitative behavior across various regions of $m_3$ can be summarized as follows:
\begin{itemize}
\item For ($m_3 \approx 0$), $m_{\beta\beta}^{\text{Std-IO}}$ lies in the range $\left( 1.53-4.92 \right)\times 10^{-2}$~eV while the light sterile contribution $|t_{14}^2 m_4|$ for $\Delta m^2_{s_2}=10^{-4}~ \left(10^{-2}\right) ~\eV^2$ is approximately $\sim  10^{-3}~\left(10^{-5}\right)$~eV which is small. In addition, other sterile contribution $|t_{15}^2 m_5|$ varies from $\left( 0.11 - 1.15 \right) \times 10^{-2}$. Therefore, the sterile contributions are not sufficient to completely cancel the standard contribution but it can reduce it up to one order of magnitude making the lowest value of $m_{\beta\beta}^{\rm SSI} \sim 10^{-3}$.

\item For the intermediate region ($m_3 \approx \sqrt{\Delta m^2_{\rm sol}}$) $m_{\beta\beta}^{\text{Std-IO}}\approx \left( 1.54-5.01 \right) \times 10^{-2}$~eV. The light sterile contribution with $\Delta m^2_{s_2}=10^{-4}$~eV$^2$ slightly increases to $|t_{14}^2 m_4| \approx (0.13-0.27)\times10^{-2}$~eV as $m_4$ becomes marginally larger. The heavier sterile term $|t_{15}^2 m_5|$ remains nearly constant at $(0.11 - 1.15 \times10^{-2})$~eV. Overall, the sterile contribution for $\mst^2 = 10^{-4}\eV^2$ is getting larger in this regime but still falls below the active neutrino contribution and hence there is no complete cancellation. For $\mst^2 = 10^{-2}\eV^2$, the active contribution is always dominant over the sterile contributions, and the domination of active neutrinos increases as the values of the lightest neutrino mass increase. Hence, for $\mst^2=10^{-2}~\eV^2$, 
the cancellation is not possible for the entire range of the lightest neutrino mass. 

\item For $0.01~\eV <m_3< 0.1~\eV$ region, it is seen that for $\mst^2 =10^{-4}\eV^2$, the sterile neutrino contributions can destructively interfere with the standard-IO contribution. Therefore, the total contribution can lie well below $10^{-4}$ eV.

\item For the quasi-degenerate region ($m_3 \gtrsim 0.2$ eV), the active neutrino contributions dominate over the sterile contributions for both values of $\mst^2$ and thus complete cancellation does not occur in this region. However, for $\mst^2 = 10^{-4}\eV^2$, the partial cancellation between the active and sterile contributions for suitable Majorana phases relaxes the limit on the lightest neutrino mass to $m_{\rm lightest}< 0.6~\eV$ as compared to the standard IO scenario,$m_{\rm lightest}< 0.3~\eV$. Consequently, this quasi-degenerate regime may yield $m_{\rm lightest}$ values exceeding $0.1$~eV, potentially in serious tension with the cosmological bounds ($\sum m_\nu < 0.16$~eV). However, in some exotic cosmological models~\cite{DiValentino:2015ola,Shao:2024mag}, the bounds on $\sum m_\nu$ can be significantly relaxed and in this scenarios \nubb~ can provide stringent limit on lightest neutrino mass.
\end{itemize}

\begin{table}[t]
\centering
\begin{tabular}{|c|c|c|c|c|}
\hline
Region & $m_{\beta\beta}^{\rm Std-IO}$ (eV)  & \multicolumn{2}{c|}{$|t_{14}^2 m_4|$ (eV)} & $|t_{15}^2 m_5|$ (eV) \\
\cline{3-4}
 & & ($\Delta m^2_{s_2}=10^{-4}\eV^2$) & ($\Delta m^2_{s_2}=10^{-2}\eV^2$) & ($\Delta m^2_{s_1}=1.3~\eV^2$) \\
\hline \hline
$m_3 \approx 0$ 
  & $\left(1.53:4.92 \right)\times 10^{-2}$
  & $\left(0.11:0.25\right)\times10^{-2}$  & $\left(0.5:5.0\right)\times10^{-5}$ 
  & $\left(0.11:1.15\right)\times10^{-2}$ \\
\hline 
$m_3 \approx \sqrt{\Delta m^2_{\rm sol}}$ 
 & $\left(1.545:5.013 \right)\times 10^{-2}$
 & $\left(0.13:0.27\right)\times10^{-2}$ 
 & $\left(0.5:5.07\right)\times10^{-4}$ 
 & $\left(0.114:1.152\right) \times 10^{-2}$ \\
\hline
$m_3 \approx 0.05~\eV$ 
 & $\left(2.11:7.05 \right)\times 10^{-2}$
 & $\left(0.57:1.27\right)\times10^{-2}$ 
 & $\left(0.56:5.6\right)\times10^{-4}$ 
 & $\left(0.114:1.152\right) \times 10^{-2}$ \\
\hline
$m_3 \approx 0.2\ \mathrm{eV}$ 
 & $\left(6.17 : 20.61 \right)\times 10^{-2}$
 & $\left(2.22:5.00\right)\times10^{-2}$ 
 & $\left(0.11:1.12\right)\times10^{-3}$ 
 & $\left(0.115:1.156\right)\times10^{-2}$ \\
\hline
\end{tabular}
\caption{Representative numerical ranges for $m_{\beta\beta}^{\text{SSI}}$ in the active-sector and sterile contributions in the $3+2$ SSI scenario.}
\label{tab:mbb-ssi}
\end{table}

\subsubsection{SNS}
\begin{figure}[t]
\centering
\includegraphics[scale=0.5]{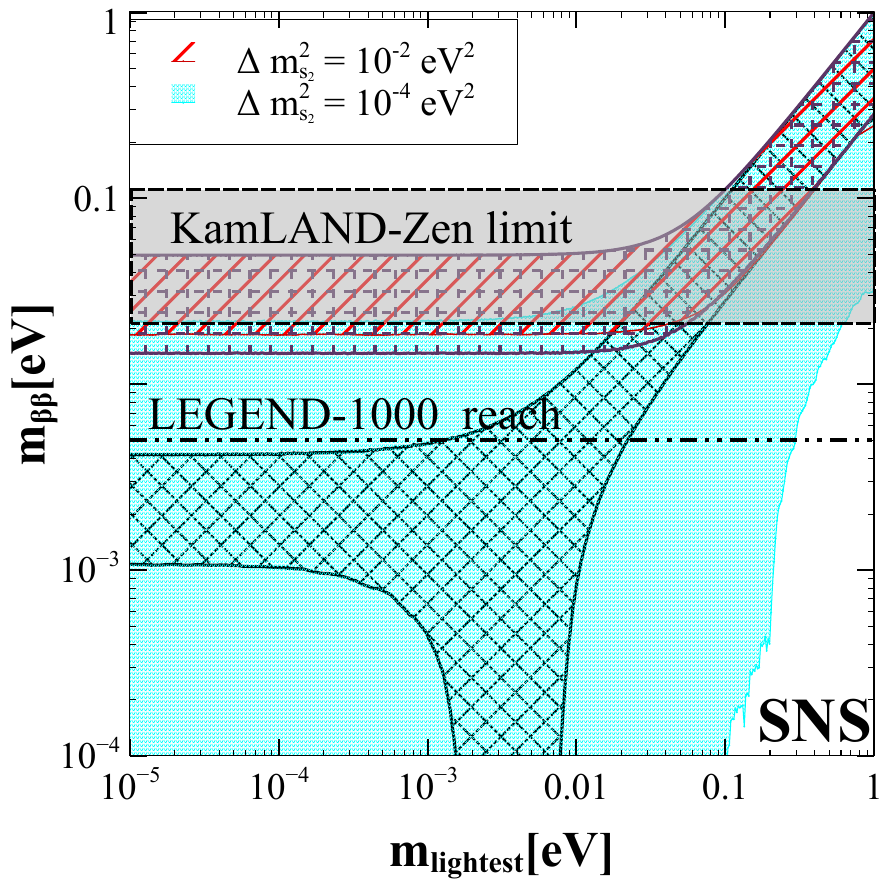}
\includegraphics[scale=0.5]{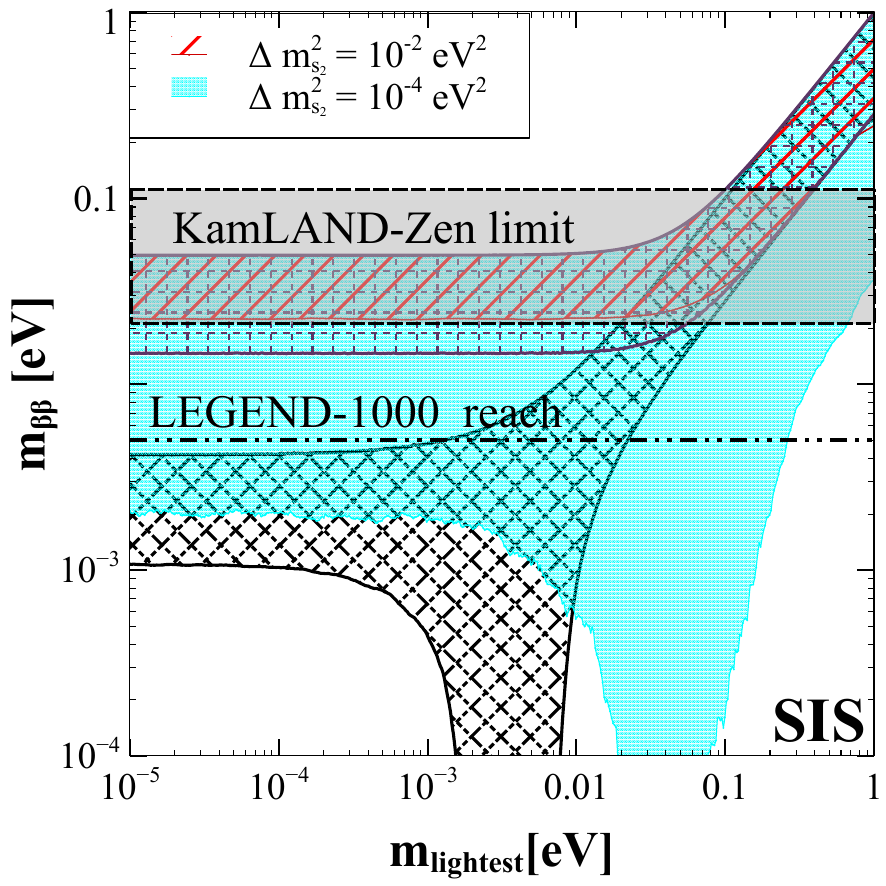}
\caption{Variation of $m_{\beta\beta}$ as a function of the light neutrino mass is shown for SNS (left panel) and SIS (right panel) scenarios. The rest of the descriptions are the same as Fig.~\ref{fig:SNS_SIS}}

\label{fig:SNS_SIS}
\end{figure}

In SNS scenario, one sterile state~($m_4$) is lighter than the other active and sterile states. In particular SNS follows the mass ordering as $m_{5}>m_{3}>m_{2}>m_{1}>m_{4}$. The relevant expression of effective Majorana mass becomes,
\begin{eqnarray}
m_{\beta\beta}^{\rm SNS} &=& c_{14}^2 c_{15}^2 \left| m_{\beta\beta}^{\rm NO} + t_{14}^2\,m_4\,e^{i\alpha_{41}} + t_{15}^2\,m_5\,e^{i\alpha_{51}} \right|,\label{eq:mbb_SNS}
\end{eqnarray} 
where $m_{\beta\beta}^{\rm NO}$ is defined in Eq.~\eqref{eq:mbb_standard} except in this case the lightest neutrino mass is $m_4$ unlike the standard NO scenario. Because of two different \(\Delta m^2_{s_2}\) values, the active neutrino mass scales is different for the two cases and hence, the values of $m_{\beta\beta}^{\rm NO}$ is different. All the relevant numerical quantities to estimate Eq.~\eqref{eq:mbb_SNS} are given in Table~\ref{tab:mbb-sns}. The left panel of Fig.~\ref{fig:SNS_SIS} shows the variation of $m_{\beta\beta}^{\rm SNS}$ as a function of the lightest neutrino state, $m_4$. From the figure, it is understood that,
\begin{itemize}
    \item For $m_4 \approx 0 $ region, one sterile neutrino contribution $\left|t_{14}^2\,m_4 \right|$ is approximately zero and the nature of $m_{\beta\beta}$ depends on the interplay of the active neutrino and the other sterile neutrino contribution. It is clearly evident from Table~\ref{tab:mbb-sns} that a complete cancellation can be possible between $m_{\beta\beta}^{\rm NO}$ and $\left| t_{15}^2\, m_5 \right|$ for $\mst^2 = 10^{-4} ~\eV^2$. It is also seen from the Fig.~\ref{fig:SSN-SSI}, that for $\mst^2 = 10^{-4} ~\eV^2$, the cancellation region extends up to $m_{\text{lightest}}\sim 0.1~\eV$ because up to this point the total sterile contribution are of similar order to the minimum value of $m_{\beta\beta}^{\text{NO}}$. Therefore, for very finely tuned parameters the cancellation occurs. However, for $\mst^2 = 10^{-2} ~\eV^2$ $m_{\beta\beta}^{\rm NO}$ dominates over the sterile contribution due to the large splitting between $m_4$ and $m_1$. This spoils the cancellation condition and $m_{\beta\beta}^{\rm SNS}$ can vary between $\left( 1.79- 11.20\right)\times 10^{-2}$ eV, some part of which is disfavored from the current KamLAND-Zen bound depending on the NMEs. Therefore, future ton-scale experiments like LEGEND-1000 can completely probe this scenario for $\mst^2 = 10^{-2}~\eV^2$.  It is also seen that for $\mst^2 = 10^{-2}~\eV^2$, the active neutrino contribution always dominates over the total sterile contribution and its get bigger for large $m_{\text{lightest}}$ values. Therefore, the cancellation remains impossible for $\mst^2 = 10^{-2}~\eV^2$. 

    \item For $\mst^2 = 10^{-4}~\eV^2$, beyond $m_{\text{lightest}} \gtrsim 0.2 ~\eV$, the total sterile contribution falls short to the total active neutrino contributions and complete cancellation is not possible. In addition, the active contribution become proportional to the $m_{\text{lightest}}$ and therefore, the minimum and maximum values becomes proportional to $m_{\text{lightest}}$ which is clearly seen in the Fig.~\ref{fig:SNS_SIS}.
\end{itemize}

\begin{table}[t]
\centering
\begin{tabular}{|c|c|c|c|c|c|}
\hline
Region & \multicolumn{2}{c|}{$m_{\beta\beta}^{\rm NO}$ (eV)} & \multicolumn{2}{c|}{$|t_{14}^2 m_4|$ (eV)}  & $|t_{15}^2 m_5|$ (eV) \\
\cline{2-5}  
 & $\Delta m^2_{s_2}=10^{-4}\eV^2$ & $\Delta m^2_{s_2}=10^{-2}\eV^2$ & $\Delta m^2_{s_2}=10^{-4}\eV^2$ & $\Delta m^2_{s_2}=10^{-2}\eV^2$ & ($\Delta m^2_{s_1}=1.3~\eV^2$) \\
\hline \hline
$m_4 \approx 0$ 
  & $\left( 0.097 : 1.21 \right) \times 10^{-2}$
  & $\left( 2.95:10.04\right)\times10^{-2}$
  & $0.0$ 
  & $0.0$ 
  & $\left( 0.11:1.15\right) \times10^{-2}$ \\
\hline
$m_4 \approx 0.1\ \mathrm{eV}$ 
 & $\left( 2.96 : 10.09 \right) \times 10^{-2}$ 
 & $\left(4.22:14.17 \right)\times 10^{-2}$
 & $\left( 1.11:2.50 \right) \times10^{-2}$ 
 & $\left( 0.5 : 5.0 \right) \times10^{-4}$ 
 & $\left( 0.11 : 1.16 \right) \times 10^{-2}$ \\
\hline
$m_4 \approx 0.3\ \mathrm{eV}$ 
 & $\left(8.95:30.03 \right)\times 10^{-2}$
 & $\left(9.40:31.64 \right)\times 10^{-2}$
 & $\left(3.33:7.50 \right) \times 10^{-2}$ 
 & $\left( 0.15 : 1.5\right) \times 10^{-3}$ 
 & $\left( 0.12 : 1.19 \right) \times10^{-2}$ \\
\hline
\end{tabular}
\caption{Representative numerical ranges for $m_{\beta\beta}^{\text{SNS}}$ in the active-sector and sterile contributions in the 3+2 SNS scenario.}
\label{tab:mbb-sns}
\end{table}

\subsubsection{SIS}
In the SIS scheme, the mass ordering of the neutrinos follows as $m_4 < m_3 < m_1 < m_2 < m_5$. The effective Majorana mass can be written as,
\begin{eqnarray}
m_{\beta\beta}^{\text{SIS}} & =& c_{14}^2 c_{15}^2\left( m_{\beta\beta}^{\text{IO}} + t_{14}^2 m_4 e^{i\alpha_{41}} + t_{15}^2 m_5 e^{i\alpha_{51}}\right)\, , \label{eq:mbb_SIS}
\end{eqnarray}
where $m_{\beta\beta}^{\text{IO}} = c_{12}^2c_{13}^2m_1 + s_{12}^2c_{13}^2m_2+ s_{13}^2 m_3$. The values are different from the standard IO case because here $m_3$ is not the lightest mass in SIS scenario, instead it is defined as $m_3 = \sqrt{m_4^2 + \mst^2}$. In the right panel of Fig.~\ref{fig:SNS_SIS}, we have shown the variation of $m_{\beta\beta}^{\text{SIS}}$ as a function of the lightest neutrino mass state~($m_4$). The relevant features of the figure is discussed as follows,
\begin{table}[H]
\centering
\begin{tabular}{|c|c|c|c|c|c|}
\hline
Region & \multicolumn{2}{c|}{$m_{\beta\beta}^{\rm IO}$ (eV)} & \multicolumn{2}{c|}{$|t_{14}^2 m_4|$ (eV)}  & $|t_{15}^2 m_5|$ (eV) \\
\cline{2-5}  
 & $\Delta m^2_{s_2}=10^{-4}\eV^2$ & $\Delta m^2_{s_2}=10^{-2}\eV^2$ & $\Delta m^2_{s_2}=10^{-4}\eV^2$ & $\Delta m^2_{s_2}=10^{-2}\eV^2$ & ($\Delta m^2_{s_1}=1.3~\eV^2$) \\
\hline \hline
$m_4 \approx 0$ 
  & $\left( 1.55 : 5.04 \right) \times 10^{-2}$
  & $\left(3.3:11.17 \right)\times 10^{-2}$
  & $0$ 
  & $0$ 
  & $\left( 0.11:1.15\right) \times10^{-2}$ \\
\hline
$m_4 \approx 0.05\ \mathrm{eV}$ 
 & $\left( 2.14 : 7.12 \right) \times 10^{-2}$ 
 & $\left( 3.65:12.24\right)\times 10^{-2}$
 & $\left( 0.56:1.25 \right) \times10^{-2}$ 
 & $\left( 0.25 : 2.5 \right) \times10^{-4}$ 
 & $\left( 0.11 : 1.15 \right) \times 10^{-2}$ \\
\hline
$m_4 \approx 0.2\ \mathrm{eV}$ 
 & $\left(6.17:20.63 \right)\times 10^{-2}$
 & $\left(6.85:22.91 \right)\times 10^{-2}$
 & $\left( 2.22:5.0 \right) \times 10^{-2}$ 
 & $\left( 0.1 : 1.0\right) \times 10^{-3}$ 
 & $\left( 0.12 : 1.17 \right) \times10^{-2}$ \\
\hline
\end{tabular}
\caption{Representative numerical ranges for $m_{\beta\beta}^{\text{SIS}}$ in the active-sector and sterile contributions in the 3+2 SIS scenario.}
\label{tab:mbb-sis}
\end{table}

\begin{itemize}
    \item For $m_{\text{lightest}}( m_4) \approx 0$, the light sterile contribution is completely negligible since $|t_{14}^2 m_4| \simeq 0$ independent of the value of $\Delta m^2_{s_2}$, as is also evident from Table~\ref{tab:mbb-sis}. Consequently, the effective Majorana mass $m_{\beta\beta}^{\text{SIS}}$ is governed by the interplay between the active neutrino sector and the heavy sterile state $m_5$. For $\Delta m^2_{s_2}=10^{-4}\,\eV^2$, the active contribution lies in the range $m_{\beta\beta}^{\rm IO} \simeq (1.55-5.04)\times10^{-2}$~eV, while the heavy sterile term $|t_{15}^2 m_5|$ is smaller, $(0.11-1.15)\times10^{-2}$~eV. Although the active term dominates, the hierarchy between the two contributions is mild enough that partial cancellations may occur for specific choices of CP phases, leading to a moderate reduction of $m_{\beta\beta}^{\text{SIS}}$. In contrast, for $\Delta m^2_{s_2}=10^{-2}\,\eV^2$, the active contribution increases to $(3.3-11.17)\times10^{-2}$~eV, whereas the sterile contribution remains unchanged. The resulting hierarchy is significantly larger, making any substantial cancellation practically impossible. As $m_4$ departs from zero, the active contribution further increases, and the disparity between active and sterile terms grows, rendering complete cancellation increasingly difficult, as reflected in Table~\ref{tab:mbb-sis}. 

    \item For $m_4 \approx 0.05~\eV$, the light sterile state contributes non-trivially to the $0\nu\beta\beta$ amplitude, particularly for $\Delta m^2_{s_2}=10^{-4}\,\eV^2$. For sizable mixing, $\sin^2\theta_{14}\sim\mathcal{O}(0.1)$, one finds $|t_{14}^2 m_4| \simeq (0.56-1.25)\times10^{-2}~\text{eV}$, which is comparable to the heavy sterile contribution and to the lower end of the active neutrino contribution, $m_{\beta\beta}^{\rm IO} \simeq (2.14$--$7.12)\times10^{-2}$~eV. As a result, the combined sterile contribution can effectively compete with the active sector, allowing for cancellations for suitable values of Majorana phases. This leads to a substantial suppression of $m_{\beta\beta}^{\text{SIS}}$, consistent with the very small values observed in Fig.~\ref{fig:SNS_SIS}. For $\Delta m^2_{s_2}=10^{-2}\,\eV^2$, however, the light sterile contribution is strongly suppressed, $|t_{14}^2 m_4|\sim10^{-4}$~eV, and becomes negligible compared to both the active and heavy sterile terms. In this case, there is no cancellation.

    \item For ($m_1 \gtrsim 0.2$ eV) region, the active neutrino contribution becomes dominant, with $m_{\beta\beta}^{\rm IO}$ rising to $(6-23)\times10^{-2}$~eV, as shown in Table~\ref{tab:mbb-sis}. Although the light sterile contribution also increases for $\Delta m^2_{s_2}=10^{-4}\,\eV^2$, it remains subdominant, while for $\Delta m^2_{s_2}=10^{-2}\,\eV^2$ it is highly suppressed. Consequently, $m_{\beta\beta}^{\text{SIS}}$ scales approximately linearly with $m_{\text{lightest}}$, and the sterile-induced cancellations become ineffective in this regime.
    We summarize the whole discussions pertaining to the implications for neutrinoless double $\beta$ decay in the $3+2$ scenario in Table~\ref{tab:remarks}.

\end{itemize}

\begin{table}[H]
\centering
\begin{tabular}{|c|c|
p{5cm}|p{4.5cm}|}
\hline
\textbf{Spectra}
&
\textbf{Region}
&
\multicolumn{2}{c|}{\textbf{Remarks}}
\\
\cline{3-4}
&  &
\textbf{(I) $\Delta m_s^2 = 10^{-4}\,\mathrm{eV}^2$}
&
\textbf{(II) $\Delta m_s^2 = 10^{-2}\,\mathrm{eV}^2$}
\\
\hline \hline
\multirow{3}{*}{SSN}
&
$m_{\rm lightest}\approx [0,0.01]\, \eV$
&
Complete cancellation can occur between total active and total sterile contributions depending on the Majorana phases.
&
same as (I)
\\
\cline{2-4}
& $m_{\rm lightest}=0.1~\mathrm{eV}$
& Complete cancellation is still viable
& The 
active neutrino contributions dominate over sterile contributions.
\\ 
\cline{2-4}
&
$m_{\rm lightest}\gtrsim 0.1~\mathrm{eV}$
& The 
active neutrino contributions dominate over sterile contributions.
&
same as (I)
\\
\hline
\multirow{3}{*}{SSI}
&
$m_{\rm lightest}\approx [0,0.01]\, \eV$
& The sterile contributions are not sufficient to completely cancel the active neutrino contribution, but can reduce up to one order of magnitude & 
same as (I)
\\
\cline{2-4}
& $0.01 ~\eV \lesssim m_{\text{lightest}} \lesssim 0.1 ~\eV$ & Complete cancellation occur between sterile and active neutrino contribution  & Active neutrino contributions still dominate over the sterile contributions. \\
\cline{2-4}
& $m_{\text{lightest}} \gtrsim 0.2\, \eV$ & Active neutrino contributions dominate over the sterile contributions.& same as (I)\\
\hline
SNS & $m_{\text{lightest}} \approx [0, 0.1]\eV$ & Complete cancellation is possible between sterile and active contribution  & Complete cancellation is not possible. The active neutrino contribution dominates.\\
\cline{2-4}
 & $m_{\text{lightest}}\gtrsim 0.2 \, \eV$ & Complete cancellation is not possible. The active neutrino contribution dominates.  & same as (I)\\
 \hline 
SIS & $m_{\text{lightest}} \approx [0, 0.02]\, \eV$ & The active contributions dominate the effective Majorana mass & same as (I) \\
\cline{2-4}
 & $ 0.02\, \eV \lesssim m_{\text{lightest}} \lesssim 0.1\, \eV$ & The maximum contribution of the sterile neutrinos can be comparable to the minimum value of the active contributions. Therefore, complete cancellation is possible for specific fine-tuned parameters.  & The active neutrino dominates over sterile contributions. \\ \cline{2-4}
 & $m_{\text{lightest}}\gtrsim 0.1\, \eV$ & The active neutrino contribution becomes dominant again & same as (I)\\
 \hline
 \end{tabular}

\caption{The table summarizes the effect of different neutrino mass spectra on the effective Majorana mass in the 3+2 scenario. }
\label{tab:remarks}
\end{table}


\section{Summary and Conclusion}\label{sec:summary}
In this work, we have performed a comprehensive study of the neutrino mass observables in the $3$ active and $2$ sterile neutrino framework. 
{In our analysis we have considered one eV-scale sterile state motivated by short-baseline anomalies and one sub-eV sterile state inspired by T2K-NO$\nu$A tensions and the solar neutrino upturn problem. 
Our primary goal is to assess the impact of the additional sterile states on the cosmological sum of neutrino masses~($\Sigma$), the effective electron neutrino mass~($m_\beta$) measured in beta-decay experiments, and the effective Majorana mass~($m_{\beta\beta}$) probed in neutrinoless double $\beta$ decay.} 

The $3+2$ neutrino spectra can be classified into four distinct mass-ordering schemes—SSN, SSI, SNS, and SIS~(see Fig.~\ref{fig:3+2spectrum})—according to the relative positioning of the active and sterile mass eigenstates. For each ordering, we derived the relevant expressions for $\Sigma$, $m_\beta$, and $m_{\beta\beta}$ and carried out a detailed numerical analysis using the latest global oscillation data, varying all parameters within their $3\sigma$ ranges and the Majorana phases between  $0$ to $\pi$. Benchmark sterile mass splittings of $\Delta m^2_{s_1}=1.3~\text{eV}^2$ and $\Delta m^2_{s_2}=10^{-2},\,10^{-4}~\text{eV}^2$ were adopted to capture both eV-scale and sub-eV sterile neutrino effects.

Cosmological observations impose particularly strong constraints on the $3+2$ scenario through the combined bounds on $N_{\rm eff}$ and $\Sigma$. Employing the ten-parameter cosmological model (10-PCM) fit, which yields $N_{\rm eff}=3.11^{+0.37}_{-0.36}$ and an upper limit $\Sigma<0.16$~eV, we find that the additional sterile states generically enhance $\Sigma^{3+2}$ relative to the standard three-flavor expectation, leading to significant tension with precision cosmology. Among the considered spectra, only the SSN ordering remains compatible with the 10-PCM bound for very small values of the lightest neutrino mass, while the SSI and SIS scenarios are entirely excluded. The SNS scenario is viable only for $\Delta m^2_{s_2}=10^{-4}\,\eV^2$ and extremely small values of the mass of the lightest neutrino, and is fully ruled out for $\Delta m^2_{s_2}=10^{-2}\,\eV^2$. These results demonstrate that current cosmological data already exclude large portions of the $3+2$ parameter space considered in this work.

For the effective electron neutrino mass~($m_\beta$), the presence of two sterile states leads to a systematic upward shift of the values of $m_\beta$, driven primarily by the eV-scale sterile contribution. In the SSN and SSI scenarios, $m_\beta$ typically lies in the range $0.03-0.13$~eV with only mild dependence on the lightest active neutrino mass, while in the SNS and SIS orderings its behavior becomes more sensitive to the sub-eV sterile mass splitting $\Delta m^2_{s_2}$. Although current bound and future sensitivities for KATRIN remain compatible with most of the allowed parameter space, the targeted reach of the Project~8 experiment can exclude the SSI and SIS scenarios entirely, while the SNS ordering can be ruled out for $\Delta m^2_{s_2}=10^{-2}\,\eV^2$. This can also probe a substantial fraction of the SSN and SNS parameter space.

Sterile neutrinos can also leave a substantial imprint on neutrinoless double $\beta$ decay through additional contributions to the decay amplitude and additional Majorana phases. We find that in the SSN and SNS scenarios, destructive interference between active and sterile contributions can suppress $m_{\beta\beta}$ down to $\mathcal{O}(10^{-4})$~eV for suitable values of the Majorana phases. It is also seen that in the SSN scenario for $\mst^2 = 10^{-4}~\eV^2$, the cancellation region becomes more extended as compared to the $\mst^2= 10^{-2}~\eV^2$. For the SSI case partial cancellations can significantly relax the lower bound on $m_{\beta\beta}$ compared to the standard inverted ordering. In contrast, the SIS scenario permits substantial suppression only in a narrow region of parameter space, with the quasi-degenerate regime remaining largely dominated by the active sector. Such lower values for the inverted schemes enable the next-generation ton-scale neutrinoless double $\beta$ decay experiments to test sterile-induced cancellations.

In conclusion, the $3+2$ sterile neutrino framework offers a rich and highly constrained phenomenology, with strong and complementary sensitivities across cosmology, $\beta$ decay, and neutrinoless double $\beta$ decay. While precision cosmological measurements already exclude significant regions of the parameter space, upcoming data from KATRIN, Project~8, and LEGEND-1000 will provide decisive tests of the remaining viable scenarios, potentially shedding light on the existence, mass ordering, and mixing structure of light sterile neutrinos.

\section*{Acknowledgment}
The authors thank Namit Mahajan for discussions. 
SG acknowledges the J.C. Bose Fellowship (JCB/2020/000011) from the Anusandhan National Research Foundation, Government of India. The work of DP, NR and SG at the Physical Research Laboratory~(PRL) was supported by the Department of Space (DoS), Government of India. HM  acknowledges INSA for supporting the work. He is indebted to PRL theory group for the hospitality. The computations were performed on the Param Vikram-1000 High Performance Computing Cluster of the PRL.

\bibliographystyle{apsrev4-2}
\bibliography{3+2sterile}

@article{KATRIN:2025lph,
    author = "Acharya, Himal and others",
    collaboration = "KATRIN",
    title = "{Sterile-neutrino search based on 259 days of KATRIN data}",
    eprint = "2503.18667",
    archivePrefix = "arXiv",
    primaryClass = "hep-ex",
    doi = "10.1038/s41586-025-09739-9",
    journal = "Nature",
    volume = "648",
    number = "8092",
    pages = "70--75",
    year = "2025"
}

@article{CMS:2022ett,
    author = "Tumasyan, Armen and others",
    collaboration = "CMS",
    title = "{Precision measurement of the Z boson invisible width in pp collisions at s=13 TeV}",
    eprint = "2206.07110",
    archivePrefix = "arXiv",
    primaryClass = "hep-ex",
    reportNumber = "CMS-SMP-18-014, CERN-EP-2022-088",
    doi = "10.1016/j.physletb.2022.137563",
    journal = "Phys. Lett. B",
    volume = "842",
    pages = "137563",
    year = "2023"
}

@article{DoubleChooz:2012gmf,
    author = "Abe, Y. and others",
    collaboration = "Double Chooz",
    title = "{Reactor electron antineutrino disappearance in the Double Chooz experiment}",
    eprint = "1207.6632",
    archivePrefix = "arXiv",
    primaryClass = "hep-ex",
    doi = "10.1103/PhysRevD.86.052008",
    journal = "Phys. Rev. D",
    volume = "86",
    pages = "052008",
    year = "2012"
}

@article{RENO:2012mkc,
    author = "Ahn, J. K. and others",
    collaboration = "RENO",
    title = "{Observation of Reactor Electron Antineutrino Disappearance in the RENO Experiment}",
    eprint = "1204.0626",
    archivePrefix = "arXiv",
    primaryClass = "hep-ex",
    doi = "10.1103/PhysRevLett.108.191802",
    journal = "Phys. Rev. Lett.",
    volume = "108",
    pages = "191802",
    year = "2012"
}

@misc{Serebrov:2021zuh,
    author = "Serebrov, A. and Samoilov, R. and Chaikovskii, M.",
    title = "{Experimental indications of the 3+1 neutrino model with one sterile neutrino}",
    eprint = "2109.12385",
    archivePrefix = "arXiv",
    primaryClass = "hep-ph",
    month = "9",
    year = "2021"
}

@misc{Cabrera:2025fmj,
    author = {Cabrera, Emilse and Jin, Miaochen and Arg{\"u}elles, Carlos A. and Esmaili, Arman},
    title = "{Searching for sub-eV Sterile Neutrinos in Neutrino Telescopes}",
    eprint = "2509.20442",
    archivePrefix = "arXiv",
    primaryClass = "hep-ph",
    month = "9",
    year = "2025"
}

@misc{Serebrov:2021ndf,
    author = "Serebrov, A. P. and Samoilov, R. M. and Chaikovskii, M. E.",
    title = "{Analysis of the result of the Neutrino-4 experiment in conjunction with other experiments on the search for sterile neutrinos within the framework of the 3 + 1 neutrino model}",
    eprint = "2112.14856",
    archivePrefix = "arXiv",
    primaryClass = "hep-ex",
    month = "12",
    year = "2021"
}

@article{Acero:2022wqg,
    author = "Acero, M. A. and others",
    title = "{White paper on light sterile neutrino searches and related phenomenology}",
    eprint = "2203.07323",
    archivePrefix = "arXiv",
    primaryClass = "hep-ex",
    reportNumber = "FERMILAB-PUB-22-318-ND-SCD-T",
    doi = "10.1088/1361-6471/ad307f",
    journal = "J. Phys. G",
    volume = "51",
    number = "12",
    pages = "120501",
    year = "2024"
}

@article{MINOS:2017cae,
    author = "Adamson, P. and others",
    collaboration = "MINOS+",
    title = "{Search for sterile neutrinos in MINOS and MINOS+ using a two-detector fit}",
    eprint = "1710.06488",
    archivePrefix = "arXiv",
    primaryClass = "hep-ex",
    reportNumber = "FERMILAB-PUB-17-430-ND",
    doi = "10.1103/PhysRevLett.122.091803",
    journal = "Phys. Rev. Lett.",
    volume = "122",
    number = "9",
    pages = "091803",
    year = "2019"
}

@article{Alekseev:2016llm,
    author = "Alekseev, I. and others",
    title = "{DANSS: Detector of the reactor AntiNeutrino based on Solid Scintillator}",
    eprint = "1606.02896",
    archivePrefix = "arXiv",
    primaryClass = "physics.ins-det",
    doi = "10.1088/1748-0221/11/11/P11011",
    journal = "JINST",
    volume = "11",
    number = "11",
    pages = "P11011",
    year = "2016"
}

@article{IceCube:2016rnb,
    author = "Aartsen, M. G. and others",
    collaboration = "IceCube",
    title = "{Searches for Sterile Neutrinos with the IceCube Detector}",
    eprint = "1605.01990",
    archivePrefix = "arXiv",
    primaryClass = "hep-ex",
    doi = "10.1103/PhysRevLett.117.071801",
    journal = "Phys. Rev. Lett.",
    volume = "117",
    number = "7",
    pages = "071801",
    year = "2016"
}

@article{NEOS:2016wee,
    author = "Ko, Y. J. and others",
    collaboration = "NEOS",
    title = "{Sterile Neutrino Search at the NEOS Experiment}",
    eprint = "1610.05134",
    archivePrefix = "arXiv",
    primaryClass = "hep-ex",
    doi = "10.1103/PhysRevLett.118.121802",
    journal = "Phys. Rev. Lett.",
    volume = "118",
    number = "12",
    pages = "121802",
    year = "2017"
}

@article{Barinov:2021asz,
    author = "Barinov, V. V. and others",
    title = "{Results from the Baksan Experiment on Sterile Transitions (BEST)}",
    eprint = "2109.11482",
    archivePrefix = "arXiv",
    primaryClass = "nucl-ex",
    doi = "10.1103/PhysRevLett.128.232501",
    journal = "Phys. Rev. Lett.",
    volume = "128",
    number = "23",
    pages = "232501",
    year = "2022"
}

@article{Abdurashitov:1996dp,
    author = "Abdurashitov, Dzh. N. and others",
    title = "{The Russian-American gallium experiment (SAGE) Cr neutrino source measurement}",
    doi = "10.1103/PhysRevLett.77.4708",
    journal = "Phys. Rev. Lett.",
    volume = "77",
    pages = "4708--4711",
    year = "1996"
}

@article{GALLEX:1997lja,
    author = "Hampel, W. and others",
    collaboration = "GALLEX",
    title = "{Final results of the Cr-51 neutrino source experiments in GALLEX}",
    reportNumber = "DAPNIA-SPP-97-26, MPI-H-V41-1997, BNL-64864, ROM2F-97-45, TUM-SFB-375-221",
    doi = "10.1016/S0370-2693(97)01562-1",
    journal = "Phys. Lett. B",
    volume = "420",
    pages = "114--126",
    year = "1998"
}

@article{MiniBooNE:2020pnu,
    author = "Aguilar-Arevalo, A. A. and others",
    collaboration = "MiniBooNE",
    title = "{Updated MiniBooNE neutrino oscillation results with increased data and new background studies}",
    eprint = "2006.16883",
    archivePrefix = "arXiv",
    primaryClass = "hep-ex",
    reportNumber = "LA-UR-20-29235, LA-UR-20-24753, FERMILAB-PUB-20-288-AD-ND",
    doi = "10.1103/PhysRevD.103.052002",
    journal = "Phys. Rev. D",
    volume = "103",
    number = "5",
    pages = "052002",
    year = "2021"
}

@article{LSND:2001aii,
    author = "Aguilar, A. and others",
    collaboration = "LSND",
    title = "{Evidence for neutrino oscillations from the observation of $\bar{\nu}_e$ appearance in a $\bar{\nu}_\mu$
 beam}",
    eprint = "hep-ex/0104049",
    archivePrefix = "arXiv",
    doi = "10.1103/PhysRevD.64.112007",
    journal = "Phys. Rev. D",
    volume = "64",
    pages = "112007",
    year = "2001"
}

@article{Weinberg:1967tq,
    author = "Weinberg, Steven",
    title = "{A Model of Leptons}",
    doi = "10.1103/PhysRevLett.19.1264",
    journal = "Phys. Rev. Lett.",
    volume = "19",
    pages = "1264--1266",
    year = "1967"
}

@article{Glashow:1961tr,
    author = "Glashow, S. L.",
    title = "{Partial Symmetries of Weak Interactions}",
    doi = "10.1016/0029-5582(61)90469-2",
    journal = "Nucl. Phys.",
    volume = "22",
    pages = "579--588",
    year = "1961"
}

@article{Gelmini:2008fq,
    author = "Gelmini, Graciela and Osoba, Efunwande and Palomares-Ruiz, Sergio and Pascoli, Silvia",
    title = "{MeV sterile neutrinos in low reheating temperature cosmological scenarios}",
    eprint = "0803.2735",
    archivePrefix = "arXiv",
    primaryClass = "astro-ph",
    doi = "10.1088/1475-7516/2008/10/029",
    journal = "JCAP",
    volume = "10",
    pages = "029",
    year = "2008"
}

@article{Archidiacono:2014nda,
    author = "Archidiacono, Maria and Hannestad, Steen and Hansen, Rasmus Sloth and Tram, Thomas",
    title = "{Cosmology with self-interacting sterile neutrinos and dark matter - A pseudoscalar model}",
    eprint = "1404.5915",
    archivePrefix = "arXiv",
    primaryClass = "astro-ph.CO",
    doi = "10.1103/PhysRevD.91.065021",
    journal = "Phys. Rev. D",
    volume = "91",
    number = "6",
    pages = "065021",
    year = "2015"
}

@article{Weinberg:1979sa,
    author = "Weinberg, Steven",
    title = "{Baryon and Lepton Nonconserving Processes}",
    reportNumber = "HUTP-79-A050",
    doi = "10.1103/PhysRevLett.43.1566",
    journal = "Phys. Rev. Lett.",
    volume = "43",
    pages = "1566--1570",
    year = "1979"
}

@article{SNO:2002tuh,
    author = "Ahmad, Q. R. and others",
    collaboration = "SNO",
    title = "{Direct evidence for neutrino flavor transformation from neutral current interactions in the Sudbury Neutrino Observatory}",
    eprint = "nucl-ex/0204008",
    archivePrefix = "arXiv",
    doi = "10.1103/PhysRevLett.89.011301",
    journal = "Phys. Rev. Lett.",
    volume = "89",
    pages = "011301",
    year = "2002"
}

@article{Super-Kamiokande:1998kpq,
    author = "Fukuda, Y. and others",
    collaboration = "Super-Kamiokande",
    title = "{Evidence for oscillation of atmospheric neutrinos}",
    eprint = "hep-ex/9807003",
    archivePrefix = "arXiv",
    reportNumber = "BU-98-17, ICRR-REPORT-422-98-18, UCI-98-8, KEK-PREPRINT-98-95, LSU-HEPA-5-98, UMD-98-003, SBHEP-98-5, TKU-PAP-98-06, TIT-HPE-98-09",
    doi = "10.1103/PhysRevLett.81.1562",
    journal = "Phys. Rev. Lett.",
    volume = "81",
    pages = "1562--1567",
    year = "1998"
}

@article{Cabrera:2025qcs,
    author = "Cabrera, Emilse and Esmaili, Arman and Nunokawa, Hiroshi and Trzeciak, Ana Maria Garcia",
    title = "{Sensitivity of Hyper-Kamiokande to sub-eV sterile neutrinos}",
    eprint = "2511.06144",
    archivePrefix = "arXiv",
    primaryClass = "hep-ph",
    doi = "10.1103/gwpz-wsgj",
    journal = "Phys. Rev. D",
    volume = "113",
    number = "1",
    pages = "015017",
    year = "2026"
}

@article{deGouvea:2022kma,
    author = "de Gouv{\^e}a, Andr{\'e} and Jusino S{\'a}nchez, Giancarlo and Kelly, Kevin J.",
    title = "{Very light sterile neutrinos at NOvA and T2K}",
    eprint = "2204.09130",
    archivePrefix = "arXiv",
    primaryClass = "hep-ph",
    reportNumber = "FERMILAB-PUB-22-247-T, CERN-TH-2022-060",
    doi = "10.1103/PhysRevD.106.055025",
    journal = "Phys. Rev. D",
    volume = "106",
    number = "5",
    pages = "055025",
    year = "2022"
}

@article{deHolanda:2003tx,
    author = "de Holanda, P. C. and Smirnov, A. Yu.",
    title = "{Homestake result, sterile neutrinos and low-energy solar neutrino experiments}",
    eprint = "hep-ph/0307266",
    archivePrefix = "arXiv",
    doi = "10.1103/PhysRevD.69.113002",
    journal = "Phys. Rev. D",
    volume = "69",
    pages = "113002",
    year = "2004"
}

@article{deHolanda:2010am,
    author = "de Holanda, P. C. and Smirnov, A. Yu.",
    title = "{Solar neutrino spectrum, sterile neutrinos and additional radiation in the Universe}",
    eprint = "1012.5627",
    archivePrefix = "arXiv",
    primaryClass = "hep-ph",
    doi = "10.1103/PhysRevD.83.113011",
    journal = "Phys. Rev. D",
    volume = "83",
    pages = "113011",
    year = "2011"
}

@article{Jana:2024xmc,
    author = "Jana, Sudip and Puetter, Lucas and Smirnov, Alexei Yu.",
    title = "{Restricting sterile neutrinos by neutrinoless double beta decay}",
    eprint = "2408.01488",
    archivePrefix = "arXiv",
    primaryClass = "hep-ph",
    doi = "10.1103/PhysRevD.111.015011",
    journal = "Phys. Rev. D",
    volume = "111",
    number = "1",
    pages = "015011",
    year = "2025"
}

@article{MicroBooNE:2025nll,
    author = "Abratenko, P. and others",
    collaboration = "MicroBooNE",
    title = "{Search for light sterile neutrinos with two neutrino beams at MicroBooNE}",
    eprint = "2512.07159",
    archivePrefix = "arXiv",
    primaryClass = "hep-ex",
    reportNumber = "FERMILAB-PUB-24-0865-PPD",
    doi = "10.1038/s41586-025-09757-7",
    journal = "Nature",
    volume = "648",
    number = "8092",
    pages = "64--69",
    year = "2025"
}

@article{KATRIN:2020dpx,
    author = "Aker, M. and others",
    collaboration = "KATRIN",
    title = "{Bound on 3+1 Active-Sterile Neutrino Mixing from the First Four-Week Science Run of KATRIN}",
    eprint = "2011.05087",
    archivePrefix = "arXiv",
    primaryClass = "hep-ex",
    doi = "10.1103/PhysRevLett.126.091803",
    journal = "Phys. Rev. Lett.",
    volume = "126",
    number = "9",
    pages = "091803",
    year = "2021"
}

@article{Adams:2020nue,
    author = {Adams, Matthew and Bezrukov, Fedor and Elvin-Poole, Jack and Evans, Justin J. and Guzowski, Pawel and Fearraigh, Br{\'\i}an {\'O}. and S{\"o}ldner-Rembold, Stefan},
    title = "{Direct comparison of sterile neutrino constraints from cosmological data, $\nu_{e}$ disappearance data and $\nu_{\mu}\rightarrow\nu_{e}$ appearance data in a $3+1$ model}",
    eprint = "2002.07762",
    archivePrefix = "arXiv",
    primaryClass = "hep-ph",
    doi = "10.1140/epjc/s10052-020-8197-y",
    journal = "Eur. Phys. J. C",
    volume = "80",
    number = "8",
    pages = "758",
    year = "2020"
}

@article{Giunti:2019fcj,
    author = "Giunti, C. and Li, Y. F. and Zhang, Y. Y.",
    title = "{KATRIN bound on 3+1 active-sterile neutrino mixing and the reactor antineutrino anomaly}",
    eprint = "1912.12956",
    archivePrefix = "arXiv",
    primaryClass = "hep-ph",
    doi = "10.1007/JHEP05(2020)061",
    journal = "JHEP",
    volume = "05",
    pages = "061",
    year = "2020"
}

@article{ANTARES:2018rtf,
    author = "Albert, A. and others",
    collaboration = "ANTARES",
    title = "{Measuring the atmospheric neutrino oscillation parameters and constraining the 3+1 neutrino model with ten years of ANTARES data}",
    eprint = "1812.08650",
    archivePrefix = "arXiv",
    primaryClass = "hep-ex",
    doi = "10.1007/JHEP06(2019)113",
    journal = "JHEP",
    volume = "06",
    pages = "113",
    year = "2019"
}

@article{Giunti:2019hkv,
    author = "Giunti, C.",
    title = "{Short-baseline neutrino oscillations with 3 + 1 non-unitary mixing}",
    eprint = "1904.02093",
    archivePrefix = "arXiv",
    primaryClass = "hep-ph",
    doi = "10.1016/j.physletb.2019.06.038",
    journal = "Phys. Lett. B",
    volume = "795",
    pages = "236--240",
    year = "2019"
}

@article{Denton:2018dqq,
    author = "Denton, Peter B. and Farzan, Yasaman and Shoemaker, Ian M.",
    title = "{Activating the fourth neutrino of the 3+1 scheme}",
    eprint = "1811.01310",
    archivePrefix = "arXiv",
    primaryClass = "hep-ph",
    doi = "10.1103/PhysRevD.99.035003",
    journal = "Phys. Rev. D",
    volume = "99",
    number = "3",
    pages = "035003",
    year = "2019"
}

@article{Gariazzo:2017fdh,
    author = "Gariazzo, S. and Giunti, C. and Laveder, M. and Li, Y. F.",
    title = "{Updated Global 3+1 Analysis of Short-BaseLine Neutrino Oscillations}",
    eprint = "1703.00860",
    archivePrefix = "arXiv",
    primaryClass = "hep-ph",
    doi = "10.1007/JHEP06(2017)135",
    journal = "JHEP",
    volume = "06",
    pages = "135",
    year = "2017"
}

@article{Gariazzo:2019gyi,
    author = "Gariazzo, S. and de Salas, P. F. and Pastor, S.",
    title = "{Thermalisation of sterile neutrinos in the early Universe in the 3+1 scheme with full mixing matrix}",
    eprint = "1905.11290",
    archivePrefix = "arXiv",
    primaryClass = "astro-ph.CO",
    doi = "10.1088/1475-7516/2019/07/014",
    journal = "JCAP",
    volume = "07",
    pages = "014",
    year = "2019"
}

@article{Archidiacono:2012ri,
    author = "Archidiacono, Maria and Fornengo, Nicolao and Giunti, Carlo and Melchiorri, Alessandro",
    title = "{Testing 3+1 and 3+2 neutrino mass models with cosmology and short baseline experiments}",
    eprint = "1207.6515",
    archivePrefix = "arXiv",
    primaryClass = "astro-ph.CO",
    doi = "10.1103/PhysRevD.86.065028",
    journal = "Phys. Rev. D",
    volume = "86",
    pages = "065028",
    year = "2012"
}

@article{Giunti:2011gz, author = "Giunti, Carlo and Laveder, Marco", title = "{3+1 and 3+2 Sterile Neutrino Fits}", eprint = "1107.1452", archivePrefix = "arXiv", primaryClass = "hep-ph", reportNumber = "EURONU-WP6-11-37", doi = "10.1103/PhysRevD.84.073008", journal = "Phys. Rev. D", volume = "84", pages = "073008", year = "2011" }

@article{Giunti:2011cp,
    author = "Giunti, Carlo and Laveder, Marco",
    title = "{Implications of 3+1 Short-Baseline Neutrino Oscillations}",
    eprint = "1111.1069",
    archivePrefix = "arXiv",
    primaryClass = "hep-ph",
    reportNumber = "EURONU-WP6-11-44",
    doi = "10.1016/j.physletb.2011.11.015",
    journal = "Phys. Lett. B",
    volume = "706",
    pages = "200--207",
    year = "2011"
}

@article{Giunti:2011hn,
    author = "Giunti, Carlo and Laveder, Marco",
    title = "{Status of 3+1 Neutrino Mixing}",
    eprint = "1109.4033",
    archivePrefix = "arXiv",
    primaryClass = "hep-ph",
    reportNumber = "EURONU-WP6-11-43",
    doi = "10.1103/PhysRevD.84.093006",
    journal = "Phys. Rev. D",
    volume = "84",
    pages = "093006",
    year = "2011"
}

@article{Goswami:2024ahm,
    author = "Goswami, Srubabati and Pachhar, Debashis and Pan, Supriya",
    title = "{Constraining the mass-spectra in the presence of a light sterile neutrino from absolute mass-related observables}",
    eprint = "2405.04176",
    archivePrefix = "arXiv",
    primaryClass = "hep-ph",
    doi = "10.1103/PhysRevD.110.015028",
    journal = "Phys. Rev. D",
    volume = "110",
    number = "1",
    pages = "015028",
    year = "2024"
}

@article{Agarwalla:2018nlx,
    author = "Agarwalla, Sanjib Kumar and Chatterjee, Sabya Sachi and Palazzo, Antonio",
    title = "{Signatures of a Light Sterile Neutrino in T2HK}",
    eprint = "1801.04855",
    archivePrefix = "arXiv",
    primaryClass = "hep-ph",
    reportNumber = "IP-BBSR-2017-16",
    doi = "10.1007/JHEP04(2018)091",
    journal = "JHEP",
    volume = "04",
    pages = "091",
    year = "2018"
}

@article{Chatterjee:2023qyr,
    author = "Chatterjee, Animesh and Goswami, Srubabati and Pan, Supriya",
    title = "{Probing mass orderings in presence of a very light sterile neutrino in a liquid argon detector}",
    eprint = "2307.12885",
    archivePrefix = "arXiv",
    primaryClass = "hep-ph",
    reportNumber = "FERMILAB-PUB-23-494-V",
    doi = "10.1016/j.nuclphysb.2023.116370",
    journal = "Nucl. Phys. B",
    volume = "996",
    pages = "116370",
    year = "2023"
}

@article{KumarAgarwalla:2019blx,
    author = "Kumar Agarwalla, Sanjib and Chatterjee, Sabya Sachi and Palazzo, Antonio",
    title = "{Physics potential of ESS$\nu$SB in the presence of a light sterile neutrino}",
    eprint = "1909.13746",
    archivePrefix = "arXiv",
    primaryClass = "hep-ph",
    reportNumber = "IP/BBSR/2019-6, IPPP/19/74",
    doi = "10.1007/JHEP12(2019)174",
    journal = "JHEP",
    volume = "12",
    pages = "174",
    year = "2019"
}

@article{K2K:2004iot,
    author = "Aliu, E. and others",
    collaboration = "K2K",
    title = "{Evidence for muon neutrino oscillation in an accelerator-based experiment}",
    eprint = "hep-ex/0411038",
    archivePrefix = "arXiv",
    doi = "10.1103/PhysRevLett.94.081802",
    journal = "Phys. Rev. Lett.",
    volume = "94",
    pages = "081802",
    year = "2005"
}

@article{T2K:2011qtm,
    author = "Abe, K. and others",
    collaboration = "T2K",
    title = "{The T2K Experiment}",
    eprint = "1106.1238",
    archivePrefix = "arXiv",
    primaryClass = "physics.ins-det",
    doi = "10.1016/j.nima.2011.06.067",
    journal = "Nucl. Instrum. Meth. A",
    volume = "659",
    pages = "106--135",
    year = "2011"
}

@article{NOvA:2021nfi,
    author = "Acero, M. A. and others",
    collaboration = "NOvA",
    title = "{Improved measurement of neutrino oscillation parameters by the NOvA experiment}",
    eprint = "2108.08219",
    archivePrefix = "arXiv",
    primaryClass = "hep-ex",
    reportNumber = "FERMILAB-PUB-21-373-ND",
    doi = "10.1103/PhysRevD.106.032004",
    journal = "Phys. Rev. D",
    volume = "106",
    number = "3",
    pages = "032004",
    year = "2022"
}

@article{MINOS:2020llm,
    author = "Adamson, P. and others",
    collaboration = "MINOS+",
    title = "{Precision Constraints for Three-Flavor Neutrino Oscillations from the Full MINOS+ and MINOS Dataset}",
    eprint = "2006.15208",
    archivePrefix = "arXiv",
    primaryClass = "hep-ex",
    reportNumber = "FERMILAB-PUB-20-253-ND",
    doi = "10.1103/PhysRevLett.125.131802",
    journal = "Phys. Rev. Lett.",
    volume = "125",
    number = "13",
    pages = "131802",
    year = "2020"
}

@article{KamLAND:2013rgu,
    author = "Gando, A. and others",
    collaboration = "KamLAND",
    title = "{Reactor On-Off Antineutrino Measurement with KamLAND}",
    eprint = "1303.4667",
    archivePrefix = "arXiv",
    primaryClass = "hep-ex",
    doi = "10.1103/PhysRevD.88.033001",
    journal = "Phys. Rev. D",
    volume = "88",
    number = "3",
    pages = "033001",
    year = "2013"
}

@article{DayaBay:2013yxg,
    author = "An, F. P. and others",
    collaboration = "Daya Bay",
    title = "{Spectral measurement of electron antineutrino oscillation amplitude and frequency at Daya Bay}",
    eprint = "1310.6732",
    archivePrefix = "arXiv",
    primaryClass = "hep-ex",
    doi = "10.1103/PhysRevLett.112.061801",
    journal = "Phys. Rev. Lett.",
    volume = "112",
    pages = "061801",
    year = "2014"
}

@article{Goswami:2007kv,
    author = "Goswami, Srubabati and Rodejohann, Werner",
    title = "{MiniBooNE results and neutrino schemes with 2 sterile neutrinos: Possible mass orderings and observables related to neutrino masses}",
    eprint = "0706.1462",
    archivePrefix = "arXiv",
    primaryClass = "hep-ph",
    doi = "10.1088/1126-6708/2007/10/073",
    journal = "JHEP",
    volume = "10",
    pages = "073",
    year = "2007"
}

@article{Goswami:2005ng,
    author = "Goswami, Srubabati and Rodejohann, Werner",
    title = "{Constraining mass spectra with sterile neutrinos from neutrinoless double beta decay, tritium beta decay and cosmology}",
    eprint = "hep-ph/0512234",
    archivePrefix = "arXiv",
    reportNumber = "TUM-HEP-615-05",
    doi = "10.1103/PhysRevD.73.113003",
    journal = "Phys. Rev. D",
    volume = "73",
    pages = "113003",
    year = "2006"
}

@article{KATRIN:2024cdt,
    author = "Aker, Max and others",
    collaboration = "KATRIN",
    title = "{Direct neutrino-mass measurement based on 259 days of KATRIN data}",
    eprint = "2406.13516",
    archivePrefix = "arXiv",
    primaryClass = "nucl-ex",
    doi = "10.1126/science.adq9592",
    journal = "Science",
    volume = "388",
    number = "6743",
    pages = "adq9592",
    year = "2025"
}

@article{Planck:2018vyg,
    author = "Aghanim, N. and others",
    collaboration = "Planck",
    title = "{Planck 2018 results. VI. Cosmological parameters}",
    eprint = "1807.06209",
    archivePrefix = "arXiv",
    primaryClass = "astro-ph.CO",
    doi = "10.1051/0004-6361/201833910",
    journal = "Astron. Astrophys.",
    volume = "641",
    pages = "A6",
    year = "2020",
    note = "[Erratum: Astron.Astrophys. 652, C4 (2021)]"
}

@article{Akita:2020szl,
    author = "Akita, Kensuke and Yamaguchi, Masahide",
    title = "{A precision calculation of relic neutrino decoupling}",
    eprint = "2005.07047",
    archivePrefix = "arXiv",
    primaryClass = "hep-ph",
    doi = "10.1088/1475-7516/2020/08/012",
    journal = "JCAP",
    volume = "08",
    pages = "012",
    year = "2020"
}

@article{Froustey:2020mcq,
    author = "Froustey, Julien and Pitrou, Cyril and Volpe, Maria Cristina",
    title = "{Neutrino decoupling including flavour oscillations and primordial nucleosynthesis}",
    eprint = "2008.01074",
    archivePrefix = "arXiv",
    primaryClass = "hep-ph",
    doi = "10.1088/1475-7516/2020/12/015",
    journal = "JCAP",
    volume = "12",
    pages = "015",
    year = "2020"
}

@article{KamLAND-Zen:2024eml,
    author = "Abe, S. and others",
    collaboration = "KamLAND-Zen",
    title = "{Search for Majorana Neutrinos with the Complete KamLAND-Zen Dataset}",
    eprint = "2406.11438",
    archivePrefix = "arXiv",
    primaryClass = "hep-ex",
    doi = "10.1103/jkf6-48j8",
    journal = "Phys. Rev. Lett.",
    volume = "135",
    number = "26",
    pages = "262501",
    year = "2025"
}

@inproceedings{Project8:2022wqh,
    author = "Esfahani, A. Ashtari and others",
    collaboration = "Project 8",
    title = "{The Project 8 Neutrino Mass Experiment}",
    booktitle = "{Snowmass 2021}",
    eprint = "2203.07349",
    archivePrefix = "arXiv",
    primaryClass = "nucl-ex",
    month = "3",
    year = "2022"
}

@article{Dasgupta:2013zpn,
    author = "Dasgupta, Basudeb and Kopp, Joachim",
    title = "{Cosmologically Safe eV-Scale Sterile Neutrinos and Improved Dark Matter Structure}",
    eprint = "1310.6337",
    archivePrefix = "arXiv",
    primaryClass = "hep-ph",
    doi = "10.1103/PhysRevLett.112.031803",
    journal = "Phys. Rev. Lett.",
    volume = "112",
    number = "3",
    pages = "031803",
    year = "2014"
}

@article{Bennett:2020zkv,
    author = "Bennett, Jack J. and Buldgen, Gilles and De Salas, Pablo F. and Drewes, Marco and Gariazzo, Stefano and Pastor, Sergio and Wong, Yvonne Y. Y.",
    title = "{Towards a precision calculation of $N_{\rm eff}$ in the Standard Model II: Neutrino decoupling in the presence of flavour oscillations and finite-temperature QED}",
    eprint = "2012.02726",
    archivePrefix = "arXiv",
    primaryClass = "hep-ph",
    reportNumber = "CPPC-2020-10",
    doi = "10.1088/1475-7516/2021/04/073",
    journal = "JCAP",
    volume = "04",
    pages = "073",
    year = "2021"
}

@article{Hannestad:2013ana,
    author = "Hannestad, Steen and Hansen, Rasmus Sloth and Tram, Thomas",
    title = "{How Self-Interactions can Reconcile Sterile Neutrinos with Cosmology}",
    eprint = "1310.5926",
    archivePrefix = "arXiv",
    primaryClass = "astro-ph.CO",
    doi = "10.1103/PhysRevLett.112.031802",
    journal = "Phys. Rev. Lett.",
    volume = "112",
    number = "3",
    pages = "031802",
    year = "2014"
}

@article{Cline:2019seo,
    author = "Cline, James M.",
    title = "{Viable secret neutrino interactions with ultralight dark matter}",
    eprint = "1908.02278",
    archivePrefix = "arXiv",
    primaryClass = "hep-ph",
    doi = "10.1016/j.physletb.2019.135182",
    journal = "Phys. Lett. B",
    volume = "802",
    pages = "135182",
    year = "2020"
}

@article{Farzan:2019yvo,
    author = "Farzan, Yasaman",
    title = "{Ultra-light scalar saving the 3 + 1 neutrino scheme from the cosmological bounds}",
    eprint = "1907.04271",
    archivePrefix = "arXiv",
    primaryClass = "hep-ph",
    doi = "10.1016/j.physletb.2019.134911",
    journal = "Phys. Lett. B",
    volume = "797",
    pages = "134911",
    year = "2019"
}

@article{Forastieri:2017oma,
    author = "Forastieri, Francesco and Lattanzi, Massimiliano and Mangano, Gianpiero and Mirizzi, Alessandro and Natoli, Paolo and Saviano, Ninetta",
    title = "{Cosmic microwave background constraints on secret interactions among sterile neutrinos}",
    eprint = "1704.00626",
    archivePrefix = "arXiv",
    primaryClass = "astro-ph.CO",
    doi = "10.1088/1475-7516/2017/07/038",
    journal = "JCAP",
    volume = "07",
    pages = "038",
    year = "2017"
}

@article{Yaguna:2007wi,
    author = "Yaguna, Carlos E.",
    title = "{Sterile neutrino production in models with low reheating temperatures}",
    eprint = "0706.0178",
    archivePrefix = "arXiv",
    primaryClass = "hep-ph",
    doi = "10.1088/1126-6708/2007/06/002",
    journal = "JHEP",
    volume = "06",
    pages = "002",
    year = "2007"
}

@article{Gelmini:2019wfp,
    author = "Gelmini, Graciela B. and Lu, Philip and Takhistov, Volodymyr",
    title = "{Cosmological Dependence of Non-resonantly Produced Sterile Neutrinos}",
    eprint = "1909.13328",
    archivePrefix = "arXiv",
    primaryClass = "hep-ph",
    doi = "10.1088/1475-7516/2019/12/047",
    journal = "JCAP",
    volume = "12",
    pages = "047",
    year = "2019"
}

@article{Gelmini:2019esj,
    author = "Gelmini, Graciela B. and Lu, Philip and Takhistov, Volodymyr",
    title = "{Visible Sterile Neutrinos as the Earliest Relic Probes of Cosmology}",
    eprint = "1909.04168",
    archivePrefix = "arXiv",
    primaryClass = "hep-ph",
    doi = "10.1016/j.physletb.2019.135113",
    journal = "Phys. Lett. B",
    volume = "800",
    pages = "135113",
    year = "2020"
}

@article{Hasegawa:2020ctq,
    author = "Hasegawa, Takuya and Hiroshima, Nagisa and Kohri, Kazunori and Hansen, Rasmus S. L. and Tram, Thomas and Hannestad, Steen",
    title = "{MeV-scale reheating temperature and cosmological production of light sterile neutrinos}",
    eprint = "2003.13302",
    archivePrefix = "arXiv",
    primaryClass = "hep-ph",
    reportNumber = "RIKEN-iTHEMS-Report-20, UT-HET 132, KEK-Cosmo-251, KEK-TH-2204,
  IPMU20-0038",
    doi = "10.1088/1475-7516/2020/08/015",
    journal = "JCAP",
    volume = "08",
    pages = "015",
    year = "2020"
}

@article{Goswami:1995yq,
    author = "Goswami, Srubabati",
    title = "{Accelerator, reactor, solar and atmospheric neutrino oscillation: Beyond three generations}",
    eprint = "hep-ph/9507212",
    archivePrefix = "arXiv",
    reportNumber = "CUPP-95-4",
    doi = "10.1103/PhysRevD.55.2931",
    journal = "Phys. Rev. D",
    volume = "55",
    pages = "2931--2949",
    year = "1997"
}

@article{Gomez-Cadenas:1995epo,
    author = "Gomez-Cadenas, J. J. and Gonzalez-Garcia, M. C.",
    title = "{Future tau-neutrino oscillation experiments and present data}",
    eprint = "hep-ph/9504246",
    archivePrefix = "arXiv",
    reportNumber = "CERN-TH-95-80, CERN-TH-95-080",
    doi = "10.1007/s002880050189",
    journal = "Z. Phys. C",
    volume = "71",
    pages = "443--454",
    year = "1996"
}

@article{Cabrera:2024rgi,
    author = "Cabrera, Emilse and Esmaili, Arman and Quiroga, Alexander A.",
    title = "{Limits on the parameter space of (3+2) sterile neutrino scenario by IceCube data}",
    eprint = "2405.10419",
    archivePrefix = "arXiv",
    primaryClass = "hep-ph",
    doi = "10.1088/1475-7516/2024/11/059",
    journal = "JCAP",
    volume = "11",
    pages = "059",
    year = "2024"
}

@article{Hagstotz:2020ukm,
    author = "Hagstotz, Steffen and de Salas, Pablo F. and Gariazzo, Stefano and Gerbino, Martina and Lattanzi, Massimiliano and Vagnozzi, Sunny and Freese, Katherine and Pastor, Sergio",
    title = "{Bounds on light sterile neutrino mass and mixing from cosmology and laboratory searches}",
    eprint = "2003.02289",
    archivePrefix = "arXiv",
    primaryClass = "astro-ph.CO",
    doi = "10.1103/PhysRevD.104.123524",
    journal = "Phys. Rev. D",
    volume = "104",
    number = "12",
    pages = "123524",
    year = "2021"
}

@article{Shao:2024mag,
    author = "Shao, Helen and Givans, Jahmour J. and Dunkley, Jo and Madhavacheril, Mathew and Qu, Frank J. and Farren, Gerrit and Sherwin, Blake",
    title = "{Cosmological limits on the neutrino mass sum for beyond-{\ensuremath{\Lambda}}CDM models}",
    eprint = "2409.02295",
    archivePrefix = "arXiv",
    primaryClass = "astro-ph.CO",
    doi = "10.1103/PhysRevD.111.083535",
    journal = "Phys. Rev. D",
    volume = "111",
    number = "8",
    pages = "083535",
    year = "2025"
}

@article{DiValentino:2015ola,
    author = "Di Valentino, Eleonora and Melchiorri, Alessandro and Silk, Joseph",
    title = "{Beyond six parameters: extending $\Lambda$CDM}",
    eprint = "1507.06646",
    archivePrefix = "arXiv",
    primaryClass = "astro-ph.CO",
    doi = "10.1103/PhysRevD.92.121302",
    journal = "Phys. Rev. D",
    volume = "92",
    number = "12",
    pages = "121302",
    year = "2015"
}

@article{Esteban:2024eli,
    author = "Esteban, Ivan and Gonzalez-Garcia, M. C. and Maltoni, Michele and Martinez-Soler, Ivan and Pinheiro, Jo{\~a}o Paulo and Schwetz, Thomas",
    title = "{NuFit-6.0: updated global analysis of three-flavor neutrino oscillations}",
    eprint = "2410.05380",
    archivePrefix = "arXiv",
    primaryClass = "hep-ph",
    reportNumber = "IFT-UAM/CSIC-24-140, YITP-SB-2024-24, IPPP/24/64, IPPP/24/64, IFT-UAM/CSIC-24-140, YITP-SB-2024-24",
    doi = "10.1007/JHEP12(2024)216",
    journal = "JHEP",
    volume = "12",
    pages = "216",
    year = "2024"
}

@misc{JUNO:2025gmd,
    author = "Abusleme, Angel and others",
    collaboration = "JUNO",
    title = "{First measurement of reactor neutrino oscillations at JUNO}",
    eprint = "2511.14593",
    archivePrefix = "arXiv",
    primaryClass = "hep-ex",
    month = "11",
    year = "2025"
}

@misc{Minakata:2025azk,
    author = "Minakata, Hisakazu",
    title = "{eV-scale sterile neutrino: A window open to non-unitarity?}",
    eprint = "2503.09280",
    archivePrefix = "arXiv",
    primaryClass = "hep-ph",
    month = "3",
    year = "2025"
}

@article{Fong:2016yyh,
    author = "Fong, Chee Sheng and Minakata, Hisakazu and Nunokawa, Hiroshi",
    title = "{A framework for testing leptonic unitarity by neutrino oscillation experiments}",
    eprint = "1609.08623",
    archivePrefix = "arXiv",
    primaryClass = "hep-ph",
    reportNumber = "YACHAY-PUB-16-02-PN",
    doi = "10.1007/JHEP02(2017)114",
    journal = "JHEP",
    volume = "02",
    pages = "114",
    year = "2017"
}

@article{Chakraborty:2019rjc,
    author = "Chakraborty, Kaustav and Goswami, Srubabati and Karmakar, Biswajit",
    title = "{Consequences of $\mu$-$\tau$ reflection symmetry for $3+1$ neutrino mixing}",
    eprint = "1904.10184",
    archivePrefix = "arXiv",
    primaryClass = "hep-ph",
    doi = "10.1103/PhysRevD.100.035017",
    journal = "Phys. Rev. D",
    volume = "100",
    number = "3",
    pages = "035017",
    year = "2019"
}

@article{Karagiorgi:2009nb,
    author = "Karagiorgi, G. and Djurcic, Z. and Conrad, J. M. and Shaevitz, M. H. and Sorel, M.",
    title = "{Viability of Delta m**2 {\textasciitilde} 1- eV**2 sterile neutrino mixing models in light of MiniBooNE electron neutrino and antineutrino data from the Booster and NuMI beamlines}",
    eprint = "0906.1997",
    archivePrefix = "arXiv",
    primaryClass = "hep-ph",
    doi = "10.1103/PhysRevD.81.039902",
    journal = "Phys. Rev. D",
    volume = "80",
    pages = "073001",
    year = "2009",
    note = "[Erratum: Phys.Rev.D 81, 039902 (2010)]"
}

@article{Donini:2001xy,
    author = "Donini, A. and Meloni, D.",
    title = "{The 2+2 and 3+1 four family neutrino mixing at the neutrino factory}",
    eprint = "hep-ph/0105089",
    archivePrefix = "arXiv",
    reportNumber = "ROMA1-TH-2001-1313",
    doi = "10.1007/s100520100777",
    journal = "Eur. Phys. J. C",
    volume = "22",
    pages = "179--186",
    year = "2001"
}

\end{document}